 \documentclass[final,5p,times,twocolumn]{elsarticle}

\usepackage{graphics}
\usepackage{color}
\usepackage{multirow}
\usepackage{latexsym}
 \usepackage{subfigure}
 \usepackage{epsfig}
\usepackage{amssymb}
\usepackage{amsthm}
\usepackage{amsmath}
\usepackage{tabularx}
\usepackage{url}
\usepackage{tikz}
\usepackage{mathrsfs}
\usepackage[subfigure]{tocloft}
\usetikzlibrary{shapes.geometric, arrows, intersections, calc}
\usepackage[doublespacing]{setspace}

\usepackage[colorlinks=true,linkcolor=blue,citecolor=red,urlcolor=blue]{hyperref}

\usepackage{multicol}
\usepackage[switch]{lineno}

\usepackage{geometry}

\tikzstyle{startstop} = [rectangle, minimum width=5cm, minimum height=1cm,text centered, draw=black]
\tikzstyle{process} = [rectangle, minimum width=5cm, minimum height=1cm, text centered, draw=black]
\tikzstyle{arrow} = [thick,->,>=stealth]

\begin{document}																													 
%
																												

\newcommand{\kvec}{\mbox{{\scriptsize {\bf k}}}}
\newcommand{\lvec}{\mbox{{\scriptsize {\bf l}}}}
\newcommand{\qvec}{\mbox{{\scriptsize {\bf q}}}}


\def\eq#1{(\ref{#1})}
\def\sec#1{\hspace{1mm}section \ref{#1}}
\def\chap#1{\hspace{1mm}section \ref{#1}}
\def\fig#1{\hspace{1mm}Fig. \ref{#1}}
\def\figur#1{\hspace{1mm}Figure \ref{#1}}
\def\tab#1{\hspace{1mm}Table \ref{#1}}

\title{Design of the wavy wall in a partially heated channel using CFD simulations and human-assisted Bayesian optimization}

\author[label1]{Piotr Kamiński}
\ead{piotr.kaminski@pcz.pl}
\author[label1]{Karol Wawrzak}
\author[label2,label3]{Yiqing Li}
\author[label2,label4]{Bernd R. Noack}
\author[label1]{Artur Tyliszczak}

\address[label1]{Department of Thermal Machinery, Czestochowa University of Technology, Al. Armii Krajowej 21, 42-200 Czestochowa, Poland}
\address[label2]{Chair of Artificial Intelligence and Aerodynamics, School of Mechanical Engineering and Automation, Harbin Institute of Technology, 518055 Shenzhen, P.R. China}
\address[label3]{Department of Mechanical Engineering, University College London, London, UK}
\address[label4]{Guangdong Provincial Key Laboratory of Intelligent Morphing Mechanisms and Adaptive Robotics, Harbin Institute of Technology, 518055 Shenzhen, P.R. China}

\begin{abstract}
This study explores heated wavy wall shape design in channel flow using machine learning, aiming to minimize temperature variation ($\sigma_T$) while limiting pressure loss ($\Delta p$). A cost function $J$ defined as a product of $\sigma_T$ and $\Delta p$ balances these competing objectives. Optimization is performed via Bayesian optimization (BO) coupled with Reynolds-Averaged Navier-Stokes (RANS) computations in an active learning loop involving up to 1000 subsequent iterations. Two shaping strategies are considered: a sinusoidal-type function defined by four parameters (two waviness amplitudes, wave count, and tilt), and a higher-dimensional approach employing a Piecewise Cubic Hermite Interpolation Polynomial (PCHIP) with 19 control points. Results show the sinusoidal design reduces $\sigma_T$ over $60$-fold but increases $\Delta p$ fourfold, while the PCHIP shape offers only a $15$-fold $\sigma_T$ reduction but with a twofold $\Delta p$ increase. Flow characteristics such as turbulent kinetic energy, pressure, temperature, and Nusselt number are examined for both optimal and suboptimal shapes along the Pareto front. The insights gained motivated a human-aided refinement of the BO result, leading to a further $17.7$\% reduction in $J$. This was achieved by replacing small-amplitude waviness periods with flat segments, which additionally significantly facilitates manufacturability.



\end{abstract}

\begin{keyword} wall corrugation \sep flow control \sep Bayesian optimization \sep CFD-RANS.
\end{keyword}

\maketitle
%
%
%

\section{Introduction} \label{intro}

Despite the continuous advancement of renewable energy sources \cite{bull2001renewable, sen2017opportunities, khaleel2022renewable, maxmut2023renewable}, reducing energy consumption remains a critical area of research within both academia and industry. Studies primarily focus on enhancing the efficiency of energy production, which still largely depends on fossil fuel combustion \cite{koyama2017role, york2019energy}, and its application across various end-user systems. Among these, heat exchangers represent a key category of devices with applications spanning the automotive \cite{witry2005thermal, krasny2016polymeric} and food industries \cite{rainieri1997convective}, as well as miniaturized cooling systems for computers \cite{ahammed2016thermoelectric, zhu2013analysis} and common household systems, including air-conditioning and heating \cite{abd2007heat, wu1997application}. The primary components of heat exchangers are channels and pipes in which the heat transfer between a flowing medium and walls occurs in thin boundary layers. In general, two types of flow control methods can be employed to enhance heat exchange across them, active and passive. Active methods involve complex systems that, while highly effective, require an external energy source. Examples include the application of electrohydrodynamic methods in which an electric field of high voltage and low current act on electrorheological fluids \cite{ohadi1991electrohydrodynamic, laohalertdecha2007review, wang2017investigation}; synthetic jets \cite{fang2013active, wang2014experimental, chandratilleke2010heat}; sprays \cite{chen2013heat, pavlova2008active}; forcing flow unsteadiness using vibrating surfaces \cite{hosseinian2018experimental, ricci2014convective}; introducing pulsating flow \cite{ye2021comprehensive, benavides2009heat, duan2022experimental}; or stirring the fluid \cite{zhang2020experimental, naik2018heat, lei2024effect}. Although the active approach is reported to be adaptable to varying flow conditions, it receives less attention due to its high cost. In contrast, passive flow control is much more attractive because it does not require any external input, which effectively improves the ecological aspect of the system. The passive approach involves introducing unsteadiness into the flow via surface shape deformation \cite{mousa2021review}, which strongly enhances mixing and, in turn, improves heat transfer. Various techniques have been employed to achieve this goal. Major examples include vortex generators \cite{fiebig1998vortices, awais2018heat}, widely used in the aerospace field; dimples \cite{turnow2012flow, rao2015heat}, commonly seen on golf balls, which can be utilized to promote mixing; various types of fins (offset, sinusoidal, louvred, and others) \cite{wang1997heat, manglik1995heat, jabardo2006experimental, SEARLE2020116030,cuevas2011thermo}; wire turbulizators \cite{yakut2004effects, akyurek2018experimental}; and twisted tapes \cite{eiamsa2009convective, garg2016heat}. These techniques have also been combined to create compound approaches, often exhibiting the primary benefits of their components \cite{kays1984compact, hesselgreaves2016compact, bhuiyan2012numerical}. These methods considerably increase turbulence intensity and in some cases may also reduce losses \cite{bergles2011recent}. Unconventional passive flow control approaches have also been examined in the literature. A prominent example is the use of porous walls, which from one side increase a heat exchange surface, and from the other, intensify a small-scale mixing process \cite{bovand2016heat, mahmood2007squeezed}.

The passive control approaches discussed thus far are highly effective in many applications. However, their intricate shapes can sometimes make them challenging to manufacture, often requiring advanced fabrication methods~\cite{Kaur_Singh_IJHMT_2021}. Consequently, simpler and more streamlined surfaces are often preferred, such as plate-fin exchangers \cite{xue2018heat} or heat sinks \cite{tan2019heat}. With advancements in additive manufacturing~\cite{SEARLE2020116030,RASTAN2020120271,Kaur_Singh_IJHMT_2021}, corrugated surfaces have garnered significant attention. These geometries disrupt boundary layers and create flow circulations, enhancing heat exchange between core and low-velocity fluid regions. An important yet well-researched example of such methods is the use of walls with grooves~\cite{saha2014heat, zheng2017turbulent, dewan2015review}. In the experimental work by Zontul et al.~\cite{zontul2021investigation}, rectangular grooves on a wall were examined, revealing that this shape significantly improved thermal performance compared to a smooth reference configuration. Moreover, the authors reported that the wall deformation promotes a more uniform distribution of vortical structures, particularly as the Reynolds number increases. Li et al.~\cite{li2008numerical} investigated heat transfer in a channel with periodically grooved rectangular walls. Their findings showed that this configuration induces flow oscillations, resulting in unsteadiness and substantial mixing enhancement. A study by Ramadhan et al. \cite{ramadhan2013groove} reported that channels with rectangular grooves generally enhance the Nusselt number ($Nu$), but this comes with the drawback of an increased pressure drop. It has been shown that employing triangular grooves instead of rectangular ones can somewhat reduce these losses. Similar observations were made by Hoang et al. \cite{hoang2021large}, where triangular corrugation was examined. Moreover, it was found that an optimal amplitude of the corrugation exists, which maximizes both $Nu$ and the so-called thermal performance factor, denoting a ratio of heat transfer rate to friction factor. Research efforts to improve heat transfer through the use of corrugations are not limited to rectangular or triangular surfaces. Other wall types, such as trapezoidal \cite{ahmed2015optimum, zhu2021fluid}, as well as combinations of rectangular, triangular, and trapezoidal shapes \cite{rabby2023convection}, have been employed with varying levels of success. These configurations generally enhance heat transfer compared to smooth walls, but their performance varies significantly depending on corrugation parameters, e.g., the depth and length of the grooves. Additionally, the shape of the grooves strongly influences pressure losses, which must be accounted for in design considerations.

A drawback of rectangular and triangular corrugation types is their sharp corners, which can lead to significant pressure losses. Additionally, sharp edges and corners cause an increase in local thermal stress, leading to premature damage to the wall in real-life scenarios. Additionally, in some extreme cases, such walls with sharp-type grooves may be difficult to manufacture with sufficient precision, particularly in small-scale devices. Due to these concerns, the use of sinusoidal walls for heat transfer enhancement is of greater interest, and thus, it has gathered widespread attention in the literature \cite{brodnianska2023heat, kumar2023numerical, madlool2023numerical, mehta2022effect, cheng2021role, dong2021study, kumari2021heat, al2020effect, alsabery2020convection, harikrishnan2019heat, dellil2004turbulent}. These works allow the formulation of the following conclusions. Sinusoidal waviness enables a considerable increase in the turbulent kinetic energy (TKE) (i); the level of the TKE increase depends on corrugation parameters, $A$ - the amplitude, $\lambda$ - the waviness period length and when $A/\lambda$ ratio grows, the heat transfer efficiency rises (ii); in some cases, however,  there is a certain threshold of $A/\lambda$ beyond which any further improvements do not occur (iii); pressure drop, considerably increasing with the growth of $A/\lambda$, is a major concern (iv). Worth noticing is that the use of wavy walls extends beyond heat transfer enhancement to the aerospace field. For instance, in aerospace science, they find applications in controlling laminar-turbulent transition~\cite{bhatia2020transition} and separation~\cite{drozdz2018passive,kaminski2024numerical}.

The summary above indicates that enhancing mixing and heat transfer in heat exchangers remains an active area of research, with wavy wall designs playing a prominent role. Optimizing these shapes for suitability in specific applications is a key focus. While some attempts have been made in this direction \cite{rostami2015optimization}, the extent of investigated wall shape modifications has been limited. The large number of potential design parameters makes the design space multidimensional and challenging to explore using traditional optimization approaches. In this context, machine learning (ML) emerges as a highly suitable approach. Its application for optimizing surface shapes in fluid mechanics dates back to the 1960s \cite{rechenberg1964kybernetische}. 
{\color{black}Our generation is witnessing a convergence of large data volumes, affordable computation, advanced algorithms, open-source resources, and industry investment in data-driven solutions. This synergy is propelling rapid advancements in machine learning, with growing applications in fluid mechanics~\cite{BurtonNoackKoumoutsakos_AnnRev_2020}. 
Machine learning frameworks such as genetic algorithms (GAs)~\cite{Benard2016ef}, genetic programming (GP)~\cite{Duriez2016book, Noack2018fssic} and reinforcement learning (RL) ~\cite{rabault2019jfm, Guastoni2023epje, nair2023jfmr} are becoming increasingly popular in flow control applications.

In the context of wavy wall shape optimization, the potential of ML remains largely untapped, with GAs being the most commonly used approach.}  
{\color{black} Fabbri~\cite{fabbri2000heat} employed GA to optimise the corrugated wall shape for maximising heat transfer. The resulting geometry featured waviness tilted in the
upstream direction with alternating wavelengths, providing a 30\% increase in transferred heat compared to a non-optimized surface. Joodaki and Ashrafizadeh~\cite{joodaki2014surface} combined GA and particle swarm optimization (PSO) and starting from a waviness tilted downstream, they obtained an optimal configuration resembling periodic cavities similar to a reverse of bump function. {\color{black}Non-dominated Sorting Genetic Algorithms (NSGA-II) is widely recognized as a classic solver for multi-objective optimization.} Wang et al.~\cite{wang2016multi} employed NSGA-II for heat transfer enhancement. While the resultant shape improved it significantly, it was accompanied by up to a $310\%$ increase in the required pumping power needed for an assumed mass flow. This was caused by substantial pressure losses induced by the waviness shape and optimization considering only the heat transfer increase. {\color{black}NSGA-II was also used by Kirkar et al.~\cite{kirkar2023multi} for designing a tube with a helically corrugated wall.} However, in this research, the design space included only two parameters, the corrugation depth and pitch. This research aimed at demonstrating a strong dependency between the control parameters and the Nusselt number.}

{GAs allow for the tuning of predefined control laws toward near-optimal performance. However, the amount of test data needed scales with the dimensionality of the parameter space, making this approach increasingly impractical for high-dimensional optimization tasks.
As the function regressor, GP and RL can provide a large model capacity for exploration. }
{However, deriving the optimal solution in finite time cannot be guaranteed}. 
{Bayesian Optimization (BO) has recently gained attention due to the capability to balance both data efficiency and the performance of the final control solution in jet and wake flows~\cite{blanchard2021ams, pino2023jfm, li2024jfm}.} 
BO is a surrogate-model-based design optimization algorithm that typically relies on Gaussian processes (GP) \cite{williams2006gaussian}. The BO procedure can be divided into two main phases. The first phase involves training a surrogate model of the objective function based on a finite training data set. In the second phase, the algorithm identifies, evaluates, and adds the most promising data samples to the data set, The surrogate model is refined, and the design space is explored through the alternation of the two phases. This approach enables fast convergence to the optimal solution with fewer evaluations compared to GA, making it particularly useful for tasks such as hyperparameter tuning in ML models or optimizing expensive simulations \cite{li2024jfm}.

{\color{black} This work focuses on the shape design of a wavy wall in a channel flow using machine learning, specifically combining BO with Computational Fluid Dynamics (CFD).} {\color{black} Regarding optimization goals, various approaches can be found in the literature. For example, Fabbri~\cite{fabbri2000heat} focused on maximizing $Nu$ while imposing a constraint on the maximum allowable pressure drop. Joodaki and Ashrafizadeh~\cite{joodaki2014surface} optimized the thermal performance factor, which included $Nu$ and friction factor, evaluated based on the pressure drop. Wang et al.~\cite{wang2016multi} performed a two-parameter optimization, where the objective functions were a pumping power and heat transfer area depended on $Nu$, friction factor, and the hydraulic diameter of the channel. 
The aim of the present research is to improve the uniformity of the temperature distribution while minimizing pressure losses by identifying a set of parameters that define the function shaping the wall. The uniform temperature of the fluid leaving a heat exchanger is advisable in various applications, e.g., in fuel preheating systems in combustion devices, cooling/heating jets impinging on walls, room ventilation systems or air curtains.} A two-parameter cost function $J$ includes temperature variations $\sigma_T$ and pressure drop $\Delta p$. Two wall-shaping functions are tested {\color{black} in the learning framework}, which differ significantly by the design parameter space. A simpler approach involves a sinusoidal type 4-parameter function, including local waviness amplitudes, number of wavinesses, and their tilt. The second, significantly more complex method, utilizes Piecewise Cubic Hermite Interpolation Polynomial (PCHIP), where the design control parameters space consists of 19 amplitudes of local function segments. The optimization-learning process employs BO working iteratively alongside the Reynolds-Averaged Navier-Stokes solver (ANSYS Fluent). A better optimal configuration is identified for the sinusoidal wall type. It reduces $\sigma_T$ by more than 60 times, with only a fourfold increase in $\Delta p$ compared to a channel with flat walls. In this regard, the use of PCHIP results in approximately a 15-fold reduction in $\sigma_T$, but with only a twofold increase in $\Delta p$. The obtained solutions, including both the optimal and the least effective configurations, are analyzed in detail to identify the reasons for their advantages and limitations. The performed investigations include comparisons of TKE, pressure, temperature, and Nusselt number distributions. The outcomes of this analysis stimulated performing a human-assisted optimization (HAO) focusing on making the manufacturing process easier without decreasing the BO-RANS achievements. Surprisingly, it turns out that HAO leads to a configuration which is significantly simpler and better from the point of $J$.

The manuscript is organized as follows. In the next section, the description of the research object, definition of the cost function and details of the wavy wall configurations are presented. Details of the optimization loop along with the description of a flow model and BO algorithm, are given in Section \ref{rans}. The obtained results, along with their in-depth analysis are presented in Section \ref{sec:results}. This section also includes the verification of the obtained RANS solution. The conclusions and suggestions for future work are provided in Section \ref{conclusions}.

\begin{figure*}[htbp!]
		\centering
        \includegraphics[angle=0, width=0.995\textwidth]{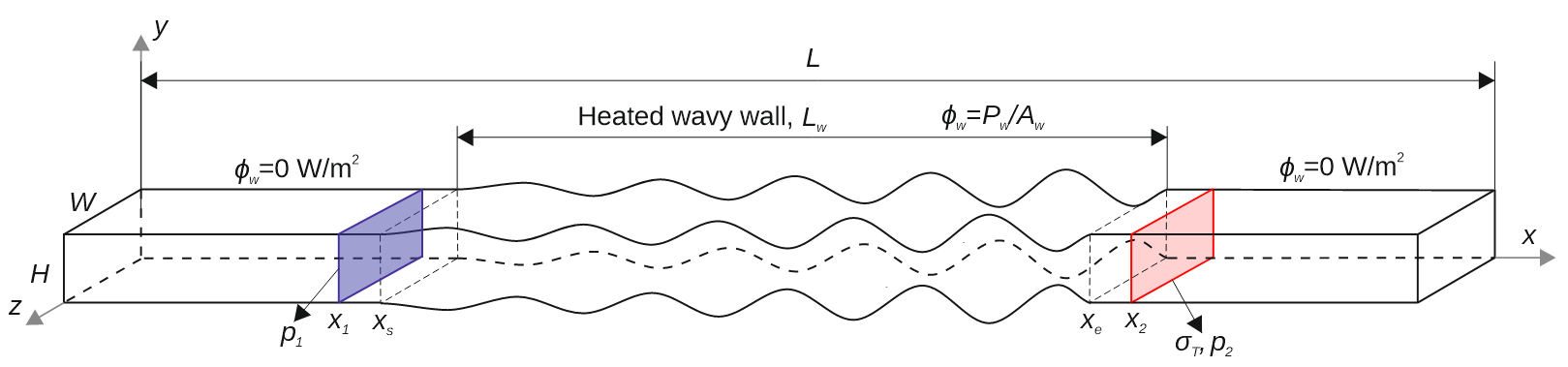}
\caption{Computational domain and the assumed boundary conditions.}
\label{fig:geometry}
\end{figure*}

\section{Research object and optimization goal\label{sec:object}}

The geometry of the channel and flow parameters have been adapted from Zontul et al.~\cite{zontul2021investigation}. Originally, it consisted of a straight inlet section followed by a series of rectangular cavities and ended with a flat wall. This configuration is analyzed in Section~\ref{sec:validation} for the numerical model validation. In the present work, we modify the central channel section by assuming a wavy wall shape, as illustrated in Fig.~\ref{fig:geometry}. {\color{black} The height ($H$), width ($W$) and length ($L$) of the channel are equal to $H=0.018$~m, $W=0.15$~m and $L=1.22$~m.} 
The wavy section has the length $L_w=0.5$~m and extends from $x_S=0.36$~m to $x_E=0.86$~m. This part of the channel is heated, and the research focuses on finding the wall shape that ensures the most uniform temperature downstream of the waviness.  

As in ~\cite{zontul2021investigation}, the inlet velocity is $U_{in}=0.1116$ m/s that results in the Reynolds number of $Re=\rho U_{in}H/\mu=2000$ ($\rho$ - the mass density, $\mu$ - the molecular viscosity). It falls well within the range of Reynolds numbers studied in the literature \cite{kurtulmucs2020heat, zontul2021investigation, liu2013comprehensive}. At the inlet, the flow is assumed to be in a low-turbulence regime with a turbulence intensity of $T_i=u'/U_{in}=1\%$, where $u'$ stands for the RMS of the velocity fluctuations. The inlet flow temperature is set at $T_{in}=300$~K. {\color{black} A relatively large dimension of the channel in the `z'-direction minimizes the influence of the side boundaries on the flow in the central region. To completely eliminate this effect, periodic boundary conditions are imposed at the side boundaries of the channel, located at $z=\pm W/2$. This approach avoids the need to account for near-corner flows, which would otherwise require specially tailored computational meshes. } All sections of the upper and lower channel walls are treated as no-slip, and their temperature varies depending on the assumed boundary conditions. In the sections before and after the waviness, the wall is treated as adiabatic with the heat flux equal to zero ($\phi_w=\kappa\nabla T=0$~W/m$^2$, $\kappa$ - the heat conductivity). In the wavy section, regardless of its shape, we assume a constant heating power equal to $P_w=10000$~W with the heat flux $\phi_w=P_w/A_w$, where $A_w$ is the surface area of the wavy wall. Note that specifying the thermal boundary conditions in terms of the heat flux implies that a local wall temperature varies based on the wall shape and flow state near the wall surface. In the case of a straight channel, the heated wall section reaches an average temperature of $T_{w,{\rm mean}}=379.2$~K. The average temperature of the fluid downstream of the heated section is independent of the wall shape and equals $T_{f,{\rm mean}}=307.9$~K.

As explained above, the objective of the optimization is to identify the waviness shape that ensures the most uniform temperature distribution downstream of the heated section. 
As a control point, we take the location right after the waviness end at $x_2=0.87$~m (see Fig.~\ref{fig:geometry}). The standard deviation quantifies the uniformity of temperature 
\begin{equation}
 \sigma_T = \frac{1}{HW} \int_0^H\int_0^W \sqrt{(T(x,y,z) - T_{f,{\rm mean}}(x))^2}\,{d}y{d}z,   
\end{equation} where $T(x,y,z)$ and $T_{f,{\rm mean}}(x)$ are the local and average temperatures at the distance $x_2$. A consequence of waving the wall shape or introducing a corrugation, e.g., by grooves \cite{elshafei2010heat}, is an increase in pressure drop along the channel. The wall shape, which enhances the mixing process and heat transfer most efficiently may lead to a very high-pressure drop ($\Delta p$) resulting in significant energy losses. In effect, the power $P=\dot{m} \Delta p$ needed to drive the flow at a given mass flow rate ($\dot{m}$) increases such that the benefits from the optimization (here $\sigma_T\to 0$) are small. Therefore, in the present work the optimization goal is to minimize $\sigma_T$ while avoiding a significant increase in the pressure drop, $\Delta p = |p_2 - p_1|$, where $p_2$ and $p_1$ are the average static pressures at locations $x_2$ and $x_1$ equal to $0.87$~m and $0.35$~m, respectively (see Fig.~\ref{fig:geometry}). By combining the parameters $\sigma_T$ and $\Delta p$, we define the cost function $J$ as follows
\begin{equation}
    J= \left(\frac{\sigma_{T}}{\sigma_{T,\rm flat}}\right)\times\left(\frac{\Delta p}{\Delta p_{\rm flat}}\right)
\end{equation}
where the variables $\sigma_{T,\rm flat}$ and $\Delta p_{\rm flat}$ stand for the reference values obtained for the channel with the flat walls. Thus, from the point of view of the optimization goal, the wall shapes leading to $J\approx 1$ can be regarded as equivalent to the flat channel, whereas the cases with $J>1$ and $J<1$ characterize worsened or improved performance, respectively. 
\textcolor{black}{This product of normalized performances in $\sigma_T$ and $\Delta p$  
weighs a relative change in each variable equally.
}{\color{black} An alternative approach could be to optimize $\sigma_T$ with a constraint imposed on $\Delta p$, as in~\cite{fabbri2000heat}, or to minimize $\sigma_T$ and $\Delta p$ separately, following the multi-objective strategy used in~\cite{wang2016multi}. The present method, in which the cost function is a single scalar variable combining $\sigma_T$ and $\Delta p$, is, however, less computationally expensive in terms of optimization time. A similar optimization strategy was employed by Joodaki and Ashrafizadeh~\cite{joodaki2014surface} to maximize the Nusselt number while minimizing friction.}

\subsection{Wall shape definition}

\begin{figure*}[h!]
		\centering
		\includegraphics[angle=0, width=0.85\textwidth]{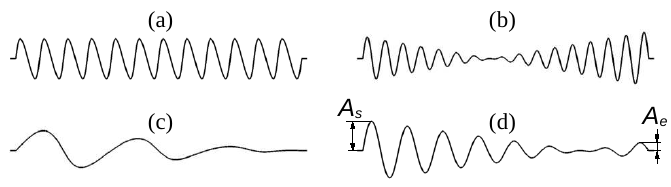}
\caption{Schematic drawings of sample wall shapes generated by the iterative procedure defined in Eq.~(\ref{eq:radii}).}
\label{fig:surface-sine}
\end{figure*}

\begin{figure*}[h!]
		\centering
            \includegraphics[angle=0, width=0.85\textwidth]{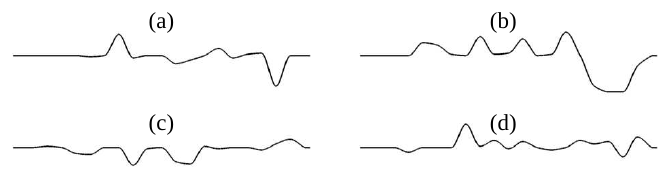}
\caption{Schematic drawings of sample wall shapes generated by PCHIP defined in Eq.~(\ref{eq:spline}).}
\label{fig:surface-spline}
\end{figure*}

We assume that the upper and lower wall shapes are symmetric with respect to the channel axis. They are defined using two methods, which differ by the number of control parameters, resulting in a smaller or greater flexibility of the waviness. The first method relies on modifying a sinusoidal wave by adjusting the number of periods (waves), their amplitudes, and tilt. Sample wall shapes obtained by applying this approach are displayed in Fig.~\ref{fig:surface-sine}. The second method employs a spline-type interpolation function (PCHIP~\cite{kahaner1989numerical}) with 19 control points. This approach allows for modifying the waviness in a free-form manner, without being restricted to the periodic pattern of the sinusoidal wave, as illustrated in Fig.~\ref{fig:surface-spline}. The following two subsections detail the applied wall-shaping methods.

\subsubsection{Sinusoidal wall type} 
In this approach, the wall shape is the result of an iterative procedure, which starts from the sinusoidal wall shape and deforms it depending on 4 control parameters. This procedure is defined as:
\begin{eqnarray}\label{eq:radii}
    y_1(x)&=&\sin \left(x\frac{2\pi N}{L_w} \right) \\
    y_{i+1}(x)&=&\sin \left(x\frac{2\pi N}{L_w} + sy_i(x)\right) + {y_i(x)};\quad i=1,...,I
\end{eqnarray}
where $y_i(x)$ represents a local amplitude of the waviness in reference to $y=\pm H/2$, $N$ denotes the number of periods along $L_w$, $s$ is a parameter that defines the tilt, and $I$ is the number of iterations $(i=1,\dots,I)$. As $I$ increases, the waviness becomes less smooth. For simplicity, we keep the number of iterations fixed and equal to $I=5$.
During the subsequent iterations, the waviness height is not constrained, meaning it can become arbitrarily large or small. In extreme cases, this could result in the channel being completely blocked if $y_i \geq H/2$. To prevent this and to control the external dimensions of the pipe, the maximum amplitude of the waviness at iteration step $I$ is normalized to one and then scaled by the desired amplitude as follows
\begin{equation}\label{eq:ampl}
    y_{I}(x)=A(x)\times \left(\frac{y_I(x)}{\max(y_I(x))}\right)
\end{equation}
with $A(x)$ varying along the streamwise direction. In the range $x\in[x_S,x_E]$ we define $A(x)$ as
\begin{equation}\label{eq:ampl2}
    A(x) = A_s + \frac{A_e - A_s}{L_w} (x-x_S) 
\end{equation}
where $A_s$ and $A_e$ are the amplitudes at the beginning and the end of the waviness, see Fig.~\ref{fig:surface-sine}. Hence, the four-parameter design space defining the wall shape consists of $A_s$, $A_e$, $N$ and $s$. 
As can be seen in Fig.~\ref{fig:surface-sine}, the shape of the wall can vary substantially, from nearly sinusoidal (a), which is obtained for $s=0$, to strongly irregular shapes (b–d). 
A limitation of this method is that it does not allow for defining a wall that is partially flat even if $A_s$ or $A_e$ is equal to zero. As can be seen in Fig.~\ref{fig:surface-sine}b-d, the amplitude of the waviness can only be equal to zero locally. 

\subsubsection{Spline-shaped wall type (PCHIP)} 

This approach does not have the aforementioned limitation. PCHIP is defined by 21 control points uniformly distributed between $x_S$ and $x_E$. Except for the first and last points, their amplitudes are treated as control parameters subject to optimization. At the first and last points, the amplitudes ($y_0$ and $y_{20}$) are set to zero. This constraint ensures the continuity of the polynomial and the channel walls and allows enforcing the condition $dy/dx=0$ at $x=x_S$ and $x=x_E$. Consequently, this approach provides a design space with 19 parameters corresponding to the amplitudes $y_{1-19}$. PCHIP in between two successive control points $x_j,\,x_{j+1}$ is defined by the following formula:

\begin{equation}\label{eq:spline}
\begin{split}
    y(x)=h_{00}y_j + &h_{10}\left(\frac{\partial y}{\partial x}\right)_j(x_{j+1}-x_j) + \\ &h_{01}y_{j+1}+h_{11}\left(\frac{\partial y}{\partial x}\right)_{j+1}(x_{j+1}-x_j)
\end{split}
\end{equation}
where $h_{00}$, $h_{10}$, $h_{01}$, $h_{11}$ are the basis functions, defined as 
\begin{eqnarray*}
    h_{00}&=&(1+2t)(1-t)^2,\,
    h_{10}=t(1-t)^2\\
    h_{01}&=&t^2(3-2t),\,
    h_{11}=t^2(t-1)
\end{eqnarray*}
with $t=(x-x_j)/h_j$, where $h_j=x_{j+1} - x_j$. To ensure the monotonicity of the polynomial, its derivatives are carefully defined, preventing the introduction of oscillations between data points. PCHIP closely follows the shape of the control points, ensuring that the polynomial extrema do not exceed those of the data points. The derivatives are defined as:

\begin{equation}
\left(\frac{\partial y}{\partial x}\right)_j =
    \begin{cases}
      0 \quad \textrm{if} \quad \Delta y_j \Delta y_{j-1} \le 0\\
       \frac{(w_1+w_2)}{(w_1/\Delta y_{j-1}+w_2/\Delta y_{j})}\quad \textrm{otherwise}
    \end{cases}
\end{equation}

with $w_1= 2h_j+h_{j-1}$, $w_2= h_j+2h_{j-1}$, $\Delta y_j =(y_{j+1}-y_j)/h_j$.


\section{Mathematical model formulation \label{rans}}

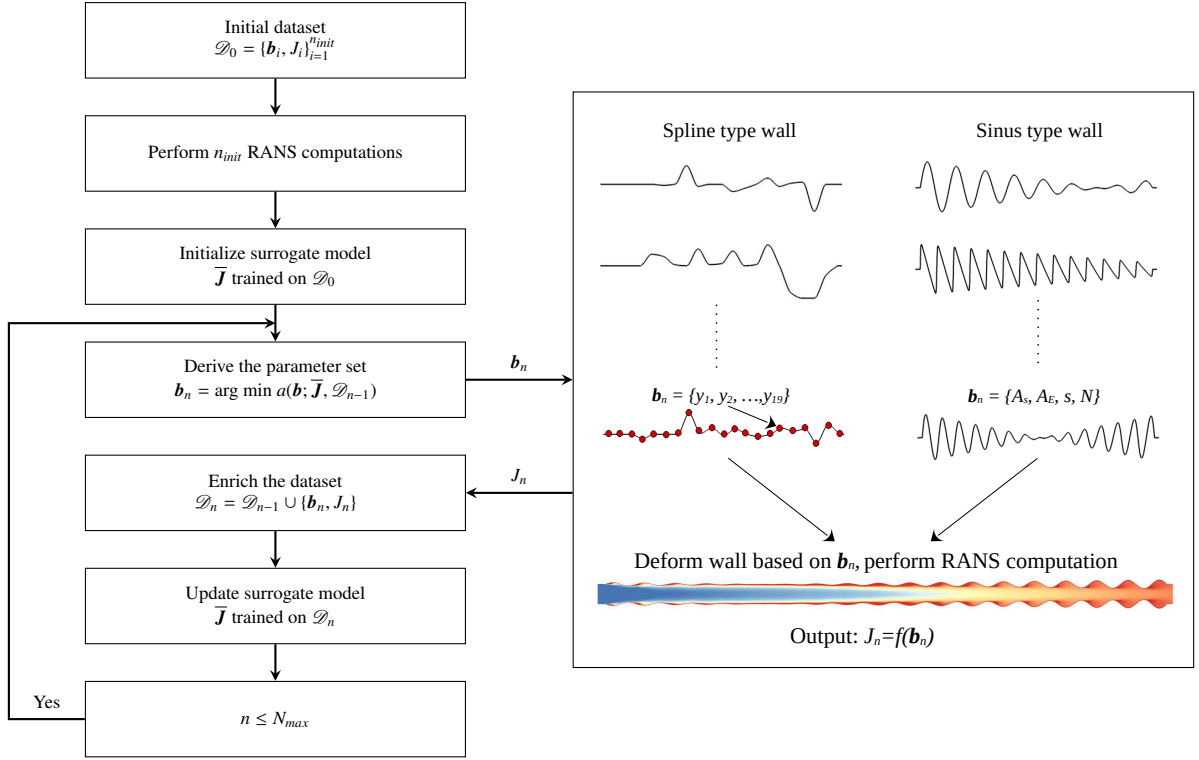
\begin{figure*}[h!]
    \centering
\begin{tikzpicture}[node distance=1.5cm]
\scriptsize
\node (start) [startstop] {\parbox{4cm}{\centering Initial dataset\\$\mathscr{D}_0 = \{\boldsymbol{b}_i, J_i \}^{n_{init}}_{i=1}$}};
\node (init-rans) [process, below of=start] {\parbox{4cm}{\centering Perform $n_{init}$ RANS computations}};
\node (initialize) [process, below of=init-rans] {\parbox{4cm}{\centering Initialize surrogate model\\ $\overline{\boldsymbol{J}}$ trained on $\mathscr{D}_0$}};
\node (derive) [process, below of=initialize] {\parbox{4cm}{\centering Derive the parameter set\\$\boldsymbol{b}_n=\text{arg min } a(\boldsymbol{b};\overline{\boldsymbol{J}},\mathscr{D}_{n-1})$}};
\node (simulate) [process, right of=derive, node distance = 8cm] {\parbox{8cm}{\centering \includegraphics[width=0.44\textwidth]{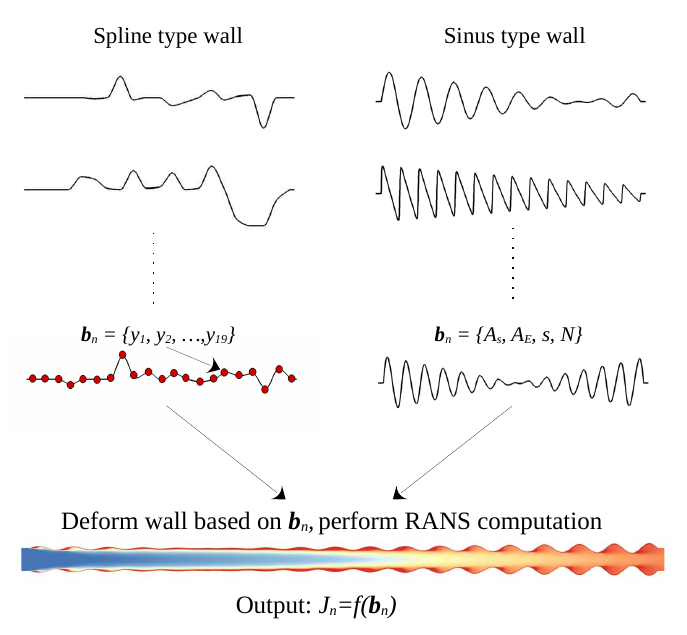}}};
\node (enrich) [process, below of=derive] {\parbox{4cm}{\centering Enrich the dataset\\$\mathscr{D}_n=\mathscr{D}_{n-1}\cup \{\boldsymbol{b}_n,J_n$\}}};
\node (update) [process, below of=enrich] {\parbox{4cm}{\centering Update surrogate model\\ $\overline{\boldsymbol{J}}$ trained on $\mathscr{D}_n$}};
\node (if) [process, below of=update] {\parbox{4cm}{\centering $n \leq N_{max}$}};

\coordinate (midArrow) at ($(initialize)!0.5!(derive)$);

\draw [arrow] (start) -- (init-rans);
\draw [arrow] (init-rans) -- (initialize);
\draw [arrow] (initialize) -- (derive);
\draw [arrow] (derive) -- node[pos=0.5, above=0.02cm] {$\boldsymbol{b}_n$} (simulate);
\draw [arrow] (simulate.west) ++(0,-1.5) -- node[pos=0.5, above=0.02cm] {$J_n$} (enrich.east);
\draw [arrow] (enrich) -- (update);
\draw [arrow] (update) -- (if);
\draw [arrow] (if.west) -- node[pos=0.5, above=0.05cm] {Yes} ++(-1,0) |- (midArrow); 
\end{tikzpicture}
\caption{Block diagram illustrating the optimization procedure.}
    \label{fig:diagram}
\end{figure*}

The optimization procedure is schematically presented in a block diagram shown in Fig.~\ref{fig:diagram}. It relies on an iterative procedure consisting of two parts working in a loop. The first one involves modeling the flow in the channel by solving the Navier-Stokes and heat transfer equations. The second is based on the Bayesian optimization method in which the solution obtained in the first step is used to compute the cost function and new wall shape parameters are defined for the next iteration. In the following two subsections, the flow model and the Bayesian optimization procedure are described, whereas details of the optimization loop are discussed in Section~\ref{results-opt}.

\subsection{Flow model}\label{sec:flow-model}
We consider an incompressible flow of water governed by the continuity, Navier-Stokes, and energy equations defined as 
\begin{equation}
      \frac{\partial \rho }{\partial t} + \frac{\partial (\rho U_i)}{\partial x_i} =0
\end{equation}

\begin{equation}
    \frac{\partial (\rho U_i)}{\partial t} +\frac{\partial (\rho U_i U_j)}{\partial x_j}   +\frac{\partial p}{\partial x_i} = \frac{\partial}{\partial x_j}\left[\mu \left( \frac{\partial U_i}{\partial x_j} + \frac{\partial U_j}{\partial x_i} - \frac{2}{3}\delta_{ij}\frac{\partial U_k}{\partial x_k}\right)\right]
\end{equation}

\begin{equation}
    \frac{\partial (\rho C_p T)}{\partial t}+ \frac{\partial (\rho C_p U_i T)}{\partial x_i}  = \frac{\partial}{\partial x_j}\left( \kappa\frac{\partial T}{\partial x_j}\right)
\end{equation}
where $U_i$ denotes the velocity component in the $i-$direction, $p$ is the pressure and $T$ represents the fluid temperature. As discussed in Section~\ref{sec:object}, the considered range of the temperature variations is small which allows assuming physical properties of water to be constant, i.e., the density equals $\rho=997$~kg/m$^3$, the molecular viscosity $\mu=0.001003$~kg/m$\,$s, the thermal conductivity $\kappa=0.6$~W/m$\,$K, and the specific heat capacity $C_p=4182$~J/kg$\,$K.

The flow at $Re=2000$ in a complex-shaped channel is treated as turbulent and is modelled by applying the Reynolds-Averaged Navier-Stokes (RANS) method. The complexity of the analyzed flow, characterized by the presence of several recirculation zones of varying sizes, precluded the unequivocal selection of a suitable RANS model. Test analysis aimed at determining the most accurate model involved a comparison of the solutions obtained from seven models, including the realizable $k-\epsilon$, Spalart-Allmaras, $k-\omega$, $k-\omega$ SST, Reynolds Stress Model (RSM), transition $k-kl-\omega$, and transition SST. Details of their formulations can be found in \cite{wilcox1998turbulence,pope2001turbulent,fluent-theory-guide}. The computations were performed by utilizing the well-known and widely used~\cite{menter1994two, saleem2023numerical, ravi2017numerical, tu2022numerical, kumar2017numerical, sharma2016computational} ANSYS Fluent commercial software. The second-order spatial discretization method was applied to the transport equations for momentum, energy, and turbulent quantities (e.g., turbulent kinetic energy). It involved using a central discretization scheme for the diffusive terms and an upwind discretization scheme for the convective terms. The coupled algorithm \cite{fluent-theory-guide} was employed for simultaneous calculations of the pressure and velocity fields, with the pressure gradient discretized using the second-order central approximation scheme.

\subsection{Bayesian optimization}

The optimization problem is to find the wall shape $\mathbf{b}^*$ minimizing the cost function $J$ defined as:
\begin{eqnarray}\label{eq:problem}
    \mathbf{b}^*=\arg\min_{\mathbf{b}\in\mathcal{B}}J(\mathbf{b}).
\end{eqnarray}
The design vector $\mathbf{b}$ consists of $n$ parameters, which vary depending on whether the sinusoidal or spline function is used for wall shaping. In the former case, $n=4$, and the design vector includes the following parameters $A_S$, $A_E$, $s$, and $N$. The amplitudes are constrained to the range $[-A_{min}, A_{max}]$ with $A_{min}=A_{max}=0.004$~m, which constitutes 22\% of the channel height. The tilt parameter varies in the range $s\in[-1,1]$ and we assume that the maximum number of periods equals $N=20$. Thus, the search space is defined as $\mathcal{B} = [-0.004, 0.004] \times [-0.004, 0.004] \times [-1, 1] \times [1, 20]$. In the case of the spline-shaped wall, $n=19$, and the design vector includes only the amplitudes of the control points, i.e., $y_{1-19}$, constrained to $[-A_{min}, A_{max}]$. {\color{black}Thus, the search space is defined as $\mathcal{B} = [-0.004, 0.004]^{19}$.}

{\color{black}
During the search for the optimal solution in Fig.~\ref{fig:diagram}, BO sequentially derives the next data point to evaluate.
The selection of data points is based on the surrogate model trained with all historical data and the acquisition function~\citep{williams2006mitbook}.

Gaussian Process (GP) regression provides a closed-form solution with posterior distribution and quantifies uncertainty, making it a standard surrogate model for BO.
Let $\mathbf{x}$ denote the input, i.e., the optimization parameter vector, and $y$ denote the corresponding output.
Based on historical data $\mathcal{D}=\{\mathbf{x}, \mathbf{y}\}$ and a Gaussian process with constant mean $m_0$, the posterior mean $\mu(\mathbf{x})$ and variance $\sigma^2(\mathbf{x})$ of the random process $\bar{f}(\mathbf{x})$ follow a normal distribution:
\begin{equation}
    \label{Eq:BO:mu}
    \mu(\mathbf{x}) = m_0 + k(\mathbf{x}, \mathbf{X})\mathbf{K}^{-1}(\mathbf{y} - m_0),
\end{equation}
\begin{equation}
    \label{Eq:BO:sigma}
    \sigma^2(\mathbf{x}) = k(\mathbf{x}, \mathbf{x}) - k(\mathbf{x}, \mathbf{X})\mathbf{K}^{-1}k(\mathbf{X},\mathbf{x}).
\end{equation}
where $\mathbf{K} = k(\mathbf{X}, \mathbf{X}) + \sigma_\epsilon^2 \mathbf{I}$.
Here, $m_0$ is the mean, and $k(\mathbf{x}, \mathbf{x})$ is the covariance function.
The covariance function uses the Radial Basis Function (RBF) kernel with relevance determination:
\begin{equation}
    k(\mathbf{x}, \mathbf{x}^\prime) = \sigma_f^2 \exp(-(\mathbf{x}-\mathbf{x}^\prime)^T \mathbf{\Theta}^{-1}(\mathbf{x}-\mathbf{x}^\prime)/2).
\end{equation}
where $\mathbf{\Theta}$ denotes hyperparameters and is calibrated to historical data using maximum likelihood estimation.
Equation \ref{Eq:BO:mu} can predict the output value at any sample $\mathbf{x}$,
and Equation \ref{Eq:BO:sigma} quantifies the uncertainty of that prediction \cite{Rasmussen2006}.

To guide the data acquisition, the likelihood-weighted confidence lower bound acquisition function \cite{blanchard2021bayesian} is used:
\begin{equation}
    \label{equ:acquisition}
    \alpha(\mathbf{x}) = \mu(\mathbf{x}) - \kappa \sigma(\mathbf{x}) w(\mathbf{x}), \quad w(\mathbf{x}) = \frac{p_{\mathbf{x}}(\mathbf{x})}{p_\mu(\mu(\mathbf{x}))}.
\end{equation} 
Here, the parameter $\kappa=1$ balances global exploration (larger $\kappa$) and local exploration (smaller $\kappa$).
The likelihood ratio $w(\mathbf{x})$ weights the input density $p_{\mathbf{x}}$ and output density $p_\mu$, measuring the uncertainty of point $\mathbf{x}$ and its expected impact on the cost function.
}

\section{Results}\label{sec:results}



In the RANS approach, the results are obtained for the time-averaged quantities~\cite{pope2001turbulent} by applying a solution algorithm involving an iterative procedure. In all analyzed cases, the initial velocity is set equal to zero, and the temperature of the fluid in the channel equals the inlet temperature. The convergent steady-state solution can be formally defined by reaching a convergence level expressed as 
\begin{equation}
\frac{\sum_{\textrm{N}_\textrm{cell}}\partial \phi/\partial t}{\sum_{\textrm{N}_\textrm{cell}}\partial \phi/\partial t\left|_{n=5}\right.}=\varepsilon,
\end{equation}
where $\phi$ denotes the velocity, temperature, and any other turbulent quantities computed with a given turbulence model (e.g., turbulent kinetic energy or turbulent dissipation rate in the $k-\epsilon$ model), ${\textrm{N}_\textrm{cell}}$ stands for the number of cells, and $n=5$ denotes the reference error level at the fifth iteration. 
{\color{black}Preliminary computations performed for several cases with significantly different wall shapes have shown that the iterative procedure converges to the prescribed threshold level $\varepsilon=10^{-6}$ within 300–600 iterations, requiring 1–2 hours on a 96 CPUs cluster. Further reduction of the error tolerance to $\varepsilon=10^{-7}$ resulted in only minor quantitative differences in the solutions.
Moreover, it has been found that the obtained results exhibit a 2D character with the maximum velocity in the `z'-direction a few orders of magnitude smaller than the wall-normal and streamwise velocity. The two-dimensional nature of the flow was confirmed by performing additional 2D simulations and comparing the results with those obtained for corresponding 3D configurations. Consequently, the RANS model validation discussed in the following section, as well as the optimization process presented later, were carried out using 2D cases. For these cases, the solutions also converged within 300–600 iterations, depending on the mesh resolution and the turbulence model applied. However, the 2D approach, which employed computational meshes with a significantly smaller number of cells (see Section~\ref{sec:2d3d}), greatly reduced the simulation time, typically to 5–10 minutes using 16 CPUs. To increase the credibility of the obtained 2D results, test computations were performed in full 3D geometries for the cases identified as the optimal and worst. As will be discussed later, the differences between the 2D and 3D simulation results are negligible.}

\subsection{Model validation \label{sec:validation}}

\begin{figure*}[h!] 
    \centering
    \includegraphics[width=0.95\linewidth]{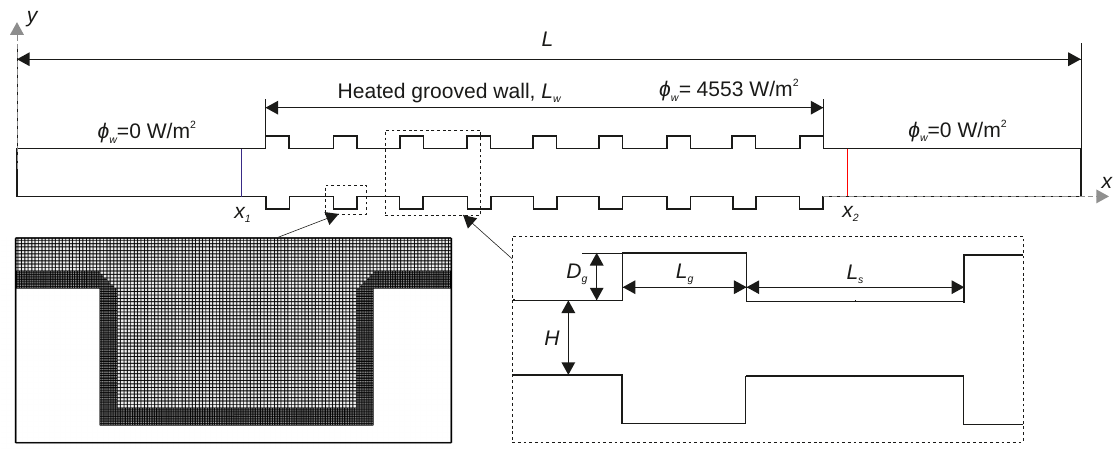}
    \caption{Schematic drawing of the experimental set-up from \cite{zontul2021investigation} with a close-up view of the mesh employed during the numerical simulations.}
    \label{fig:exp}
\end{figure*}

In this section, we analyze the accuracy level of the applied RANS method and select the model to be used for the BO-RANS procedure. As a test case, we take a flow in a channel presented in Fig.~\ref{fig:exp} studied by Zontul et al. \cite{zontul2021investigation}, which also serves as the basic configuration for the present studies. The length and width of the channel and the beginning and length of the heated section are the same as in the configuration subjected to the optimization, see Section~\ref{sec:object}. The main difference is the shape of the heated part, composed of rectangular grooves. Each groove has a length of $L_g=0.02$ m and a depth of $D_g=0.5L_g$. A distance of $L_s=2L_g$ spaces apart the grooves. The initial and end parts of the channel are insulated ($\phi_w=0$~W/m$^2$), while on the heated part $\phi_w=4553$~W/m$^2$. The inlet flow temperature is slightly lower than assumed for the present studies and equals 293~K. However, the Reynolds number is the same ($Re=2000$), which means that the flow and temperature distributions are expected to characterize a similar complexity as in the channel with walls defined by sinusoidal or spline shapes analyzed later in Section~\ref{bo-rans-sim}. 

\begin{figure}[h!] 
    \centering
    {\includegraphics[width=0.49\textwidth]{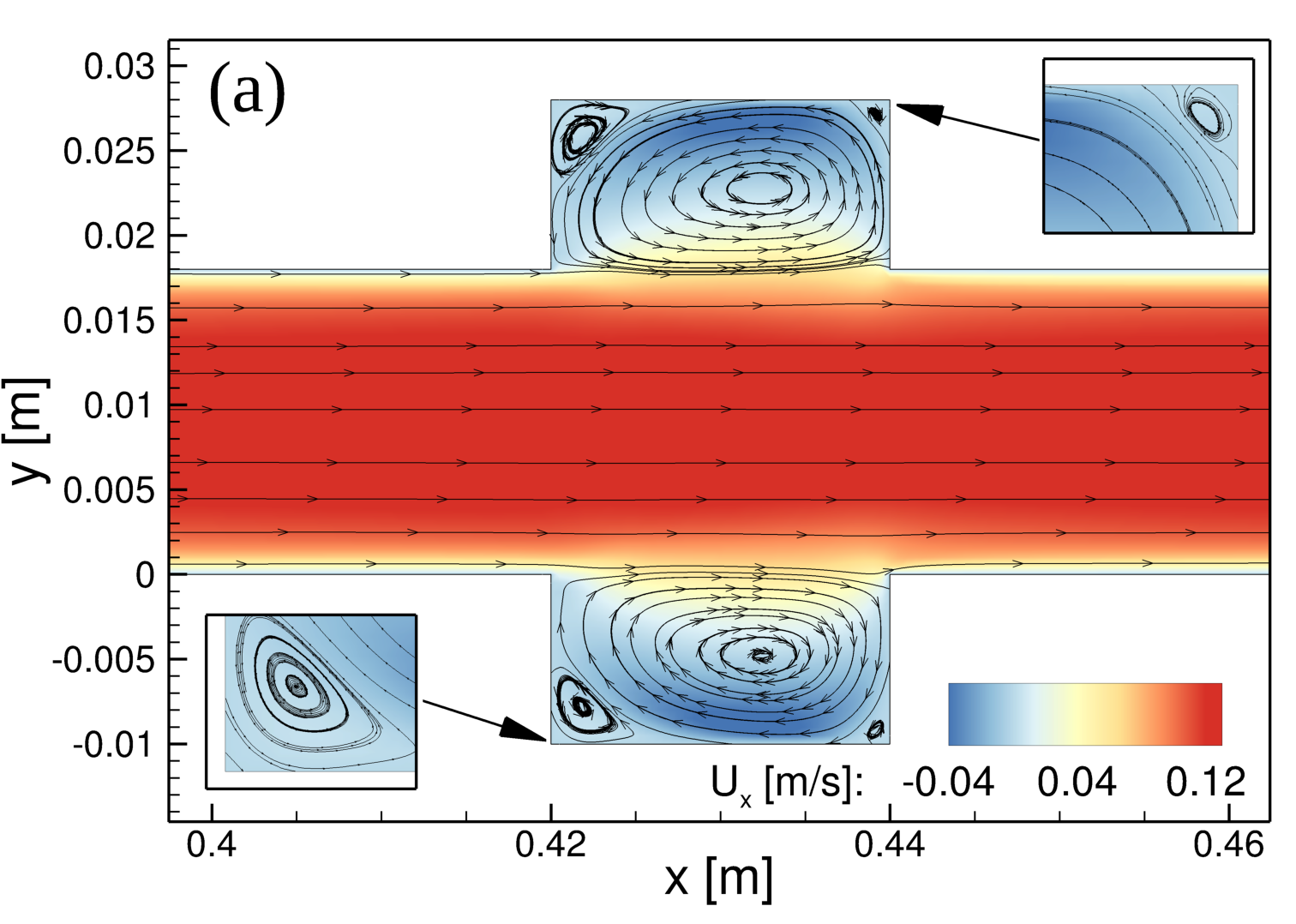}}\\
    {\includegraphics[width=0.49\textwidth]{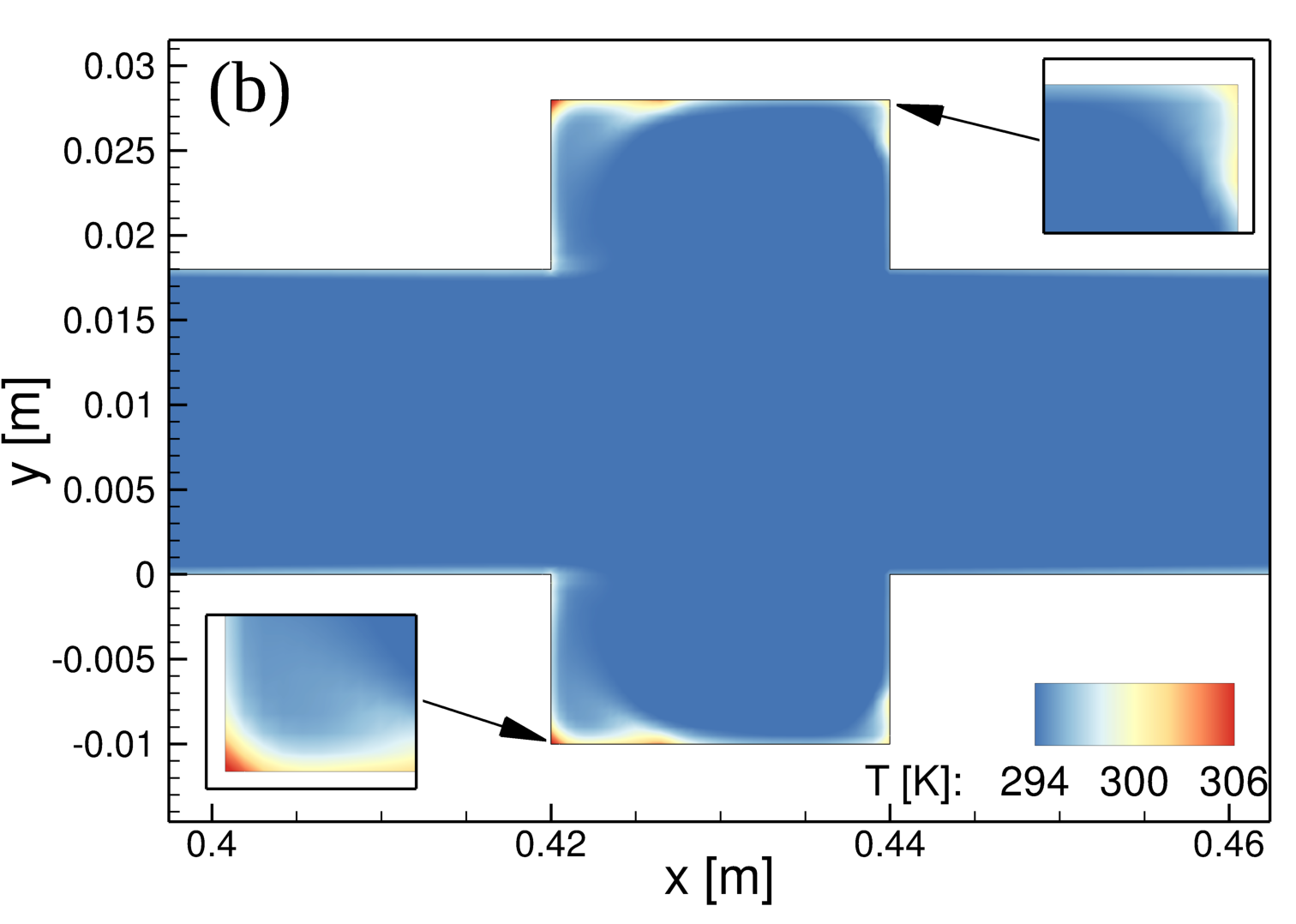}}
    \caption{Contour maps of the streamwise velocity (a) and temperature (b) within the 2nd groove and its vicinity.}
    \label{fig:recirculations-grooves}
\end{figure}

The computational mesh consists of 575340 rectangular cells. As presented in Fig.~\ref{fig:exp}, the cells are compacted near the walls to improve the resolution of the near-wall flow in the regions of large temperature gradients. The simulations are performed by applying all RANS models mentioned in Section~\ref{sec:flow-model}. Sample results obtained with the $k-\epsilon$ model are presented in Fig.~\ref{fig:recirculations-grooves}. It shows the contours of the streamwise velocity ($U_x$) and temperature inside and near the second groove. We note that, qualitatively, the presented solutions can be regarded as representative of the remaining grooves, in which the flow exhibits very similar features. Streamlines shown in Fig.~\ref{fig:recirculations-grooves}a reveal the presence of recirculation zones, within which the local streamwise velocity takes negative values. This promotes extensive mixing, intensifying the heat transfer between the wall and the flow. It can be seen that a visible increase in temperature is present only in the corners of the grooves, where small recirculation zones are formed. Along the straight channel walls, the thermal layer is very thin, and in effect, in the central flow part, the temperature is almost uniform. Its value does not diverge significantly from the inlet temperature. Downstream of the heated grooved section, in the location $x_2$, the average temperature equals $293.9$~K.

\begin{figure*}[h!] 
    \centering
    \includegraphics[width=0.99\linewidth]{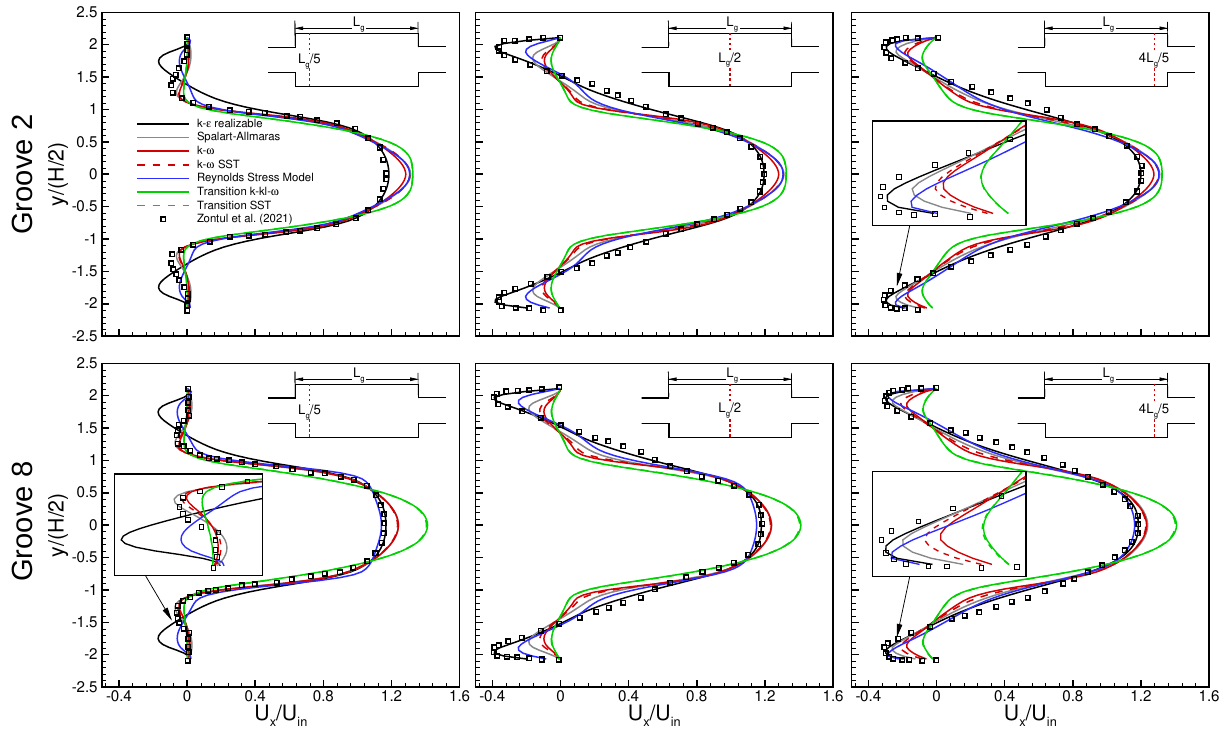}
    \caption{Profiles of the streamwise velocity obtained by applying different RANS models compared to experimental data of Zontul et al. \cite{zontul2021investigation}.}
    \label{fig:model-validation}
\end{figure*}

Figure~\ref{fig:model-validation} shows the profiles of the streamwise velocity normalized by the inlet velocity ($U_x/U_{in}$) in three cross-sections ($L_g/5,\, L_g/2,\,4L_g/5$) of the second and eighth grooves. The color lines refer to the RANS solutions obtained using different models, and the square symbols denote the experimental data from \cite{zontul2021investigation}. It can be seen that in some cases the solutions are hardly distinguishable, e.g., the profiles obtained by applying the $k-\omega$ and $k-\omega-SST$ models or those predicted with the use of transition-$k-kl-\omega$ and transition-$SST$ models nearly overlap, especially in the region $-1<y/(H/2)<1$. Regarding the accuracy of particular models, their superiority largely depends on the groove number and location where the comparison is made. For instance, in the 8th groove, in the location $L_g/5$ for $y/(H/2)<-1$ and $y/(H/2)>1$, the recirculation zones where $U_x/U_{in}<0$ are very well captured by the $k-\omega$ and transition type models. On the other hand, in the central flow part, the solutions obtained by using these models diverge from the measured values. The computed values are significantly larger and this tendency is also found in the case of using the RSM model in the second groove. Among all the presented solutions, the ones obtained by applying the $k-\epsilon$ model are the most accurate. In this case, the discrepancies are seen only in the location $L_g/5$ close to the wall of the groove. In the locations $L_g/2$ and $4L_g/5$, the solutions agree almost perfectly with the experimental data. Hence, the $k-\epsilon$ model is selected for the BO-RANS optimization procedure. It is believed that its demonstrated reliability will also apply to channels with wavy walls, where, due to the absence of sharp corners, obtaining accurate results should theoretically be simpler.

\subsection{BO-RANS simulation parameters \label{results-opt}}

\subsubsection{BO initialization}

Referring to Fig.~\ref{fig:diagram}, {\color{black}the optimization process begins with $n_{init}$ initial wall shapes, each defined by a set of parameters $\mathbf{b}$ selected via Latin hypercube sampling.} For the sinusoidal wall shape, $n_{init}=50$, while for the spline-type wall, $n_{init}=100$. In this case, the use of the larger set of initial solutions is motivated by the richer space of the control parameters when using the spline function for wall shaping (4 control parameters for the sinusoidal wall type, 19 control parameters for the spline-shaped wall). The obtained initial solutions with the associated values of the cost function $J$ are used to initialize a surrogate model. Subsequently, the main loop of the BO method starts and continues until the maximum number of iterations ($N_{max}$) is reached. 
We take $N_{max}=900$ and $N_{max}=1300$ for sinusoidal and spline-shaped wall-type,  respectively. {\color{black} Following the same reasoning behind selecting different values of $n_{\text{init}}$, the choice of a larger $N_{\text{max}}$ for the spline-shaped wall was motivated by the broader parameter space and the anticipated longer optimization process. As will be shown later, the selected $N_{\text{max}}$ values were sufficiently large to yield only slightly different learning curves for both wall types. Further increasing $N_{\text{max}}$ does not affect the optimization outcomes but leads to an unjustified increase in computational cost. Moreover, as noted by Blanchard and Sapsis~\cite{blanchard2021bayesian}, an excessively large number of iterations undermines the fundamental purpose of the BO algorithm. }
During the optimization process, the subsequent RANS solutions are stored in external files for postprocessing and controlling the computational mesh quality (see Section~\ref{sec:mesh}), while the corresponding $J_m$ and $\mathbf{b}_m$ parameters are utilized in the learning and optimization process. 

\subsubsection{Geometry and mesh generation procedure}\label{sec:mesh}

\begin{figure}[h] 
    \includegraphics[width=0.975\linewidth]{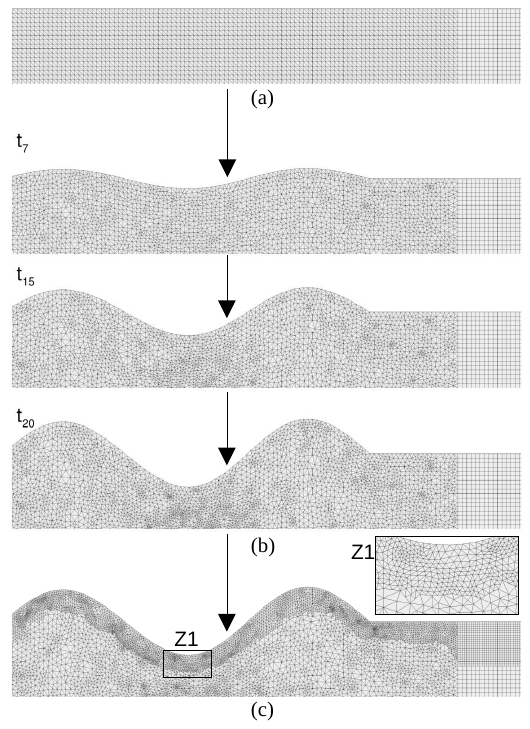}
    \caption{Three-stage wall shaping method and mesh generation procedure: (a) - the initial channel/mesh configuration; (b) - intermediate steps; (c) - the final channel/mesh configuration with the near-wall mesh refinement.}
    \label{fig:mesh}
\end{figure}

During the initialization of the BO-RANS procedure and in the main loop, the set of parameters $b^*_{m}$ changes, and the channel walls are shaped according to Eq.~(\ref{eq:radii}) or Eq.~(\ref{eq:spline}). This step is performed by applying an iterative approach in which both the wall shape and the computational mesh are successively modified. As an example illustrating this process, we present a test case with the wavy wall defined by Eq.~(\ref{eq:radii}). 
The initial configuration is a straight channel, with the computational mesh shown in Fig.~\ref{fig:mesh}a. In the region where the channel wall has to be shaped, the mesh is unstructured and consists of triangular elements. In the channel sections preceding and downstream of the waviness, the mesh is structured and composed of quadrilateral cells. In the next steps, the wall is successively deformed and the mesh adapts to the new geometry employing a dynamic meshing adaptation method available in ANSYS Fluent software \cite{fluent-theory-guide}. It involves smoothing and re-meshing algorithms, which change the mesh incrementally. To avoid overstretching of individual cells and to ensure a high quality of the mesh, e.g., low skewness, aspect ratio close to one, the wall deformation process is divided into 20 iterations. This number has been found adequate in a trial-and-error procedure relying on the generation of test meshes for various complex wall shapes. Figure~\ref{fig:mesh} shows intermediate channel shapes and meshes at the iterations $t_7$, $t_{15}$ and $t_{20}$ for the case with the wavy wall defined by the parameters $A_1=-0.0016$ m, $A_2=0.004$ m, $s=-0.02$, $N=18$ in Eq.~(\ref{eq:radii}). It can be seen in Fig.~\ref{fig:mesh}b that at $t_{20}$ the density of the mesh is not uniform and the sizes of the cells locally change. However, none of them is exaggeratedly small/large or skewed.  At this stage, the cells closest to the wall are refined by dividing them into smaller ones, creating 10 additional layers as shown in Fig.~\ref{fig:mesh}c. Note that the mesh refining process is applied along the entire channel walls including its wavy and straight parts. This ensures that the near-wall viscous layer is well-resolved also in the location $x_2$ (see Fig.~\ref{fig:geometry}), where the cost function $J$ is calculated. The near-wall mesh resolution is verified by checking the values of $y^+ = u_\tau h_w / \nu$ ($u_\tau$ - the friction velocity, $h_w$ - the height of the computational cell adjacent to the wall) at the end of every simulation for all analyzed cases. It is found that $y^+ \leq 1$, which is commonly treated as the necessary condition for applying the RANS method without involving wall model functions \cite{pope2001turbulent}. {\color{black} Depending on the shape of the waviness, the number of computational cells varies only slightly, $3.3\times 10^5$ - $3.5\times 10^5$, which practically does not impact the memory requirement and load balancing in parallel simulations. However, as discussed at the beginning of this section, reaching the convergent solution changes depending on the wall configuration and requires 300-600 iterations. These differences originate from different flow complexity in particular cases.}




\subsection{BO-RANS simulation results \label{bo-rans-sim}}

The optimization procedure is preceded by test computations for the channel with flat walls and the channel with the geometry of the walls used to validate the RANS models in Section~\ref{sec:validation}. In the latter case, to ensure consistency with the boundary conditions assumed in the optimization process, the grooved wall is heated with a heat power of $P_w = 10000$~W. The results show a pressure drop of $\Delta p = 16.039$~Pa and a temperature variance of $\sigma_T = 1.120$~K, yielding $J = 0.205$. In comparison to the flat channel, where $J = 1$, $\Delta p_{flat} = 7.919$~Pa, and $\sigma_{T,flat} = 11.086$~K, the temperature uniformity is significantly improved, though this is accompanied by more than a doubling of the pressure drop. 

\begin{figure*}[h!] 
    \centering
    \includegraphics[width=0.99\linewidth]{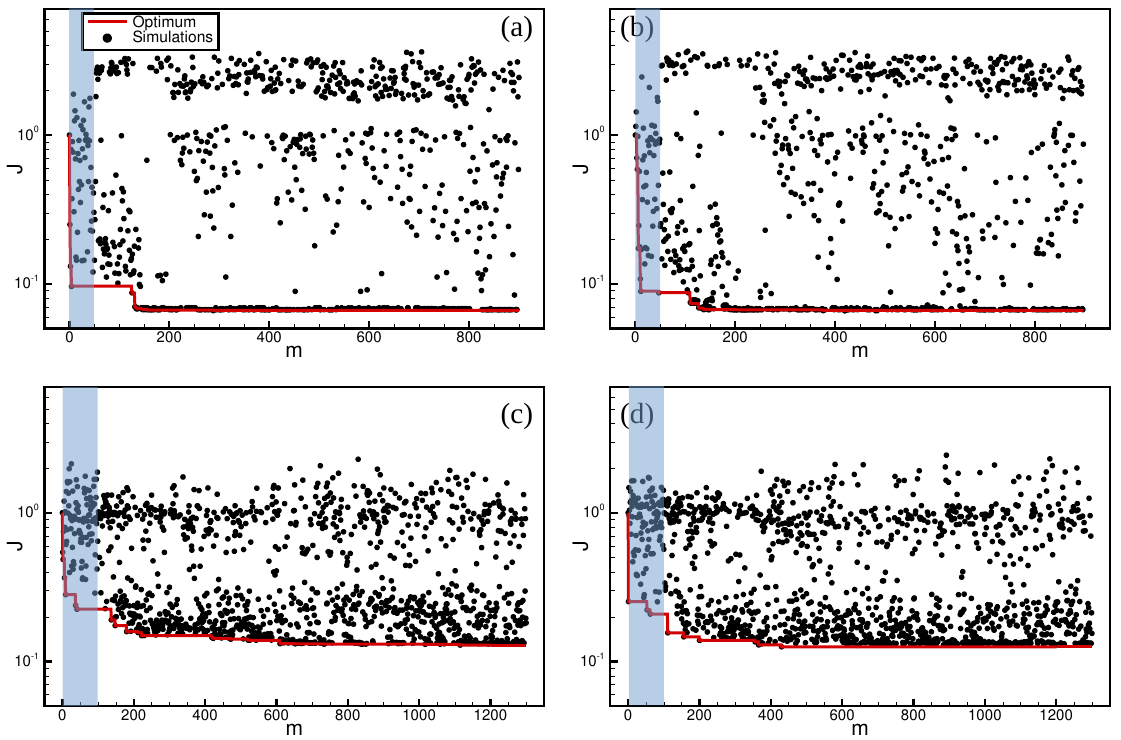}
    \caption{Learning curves of $J$ obtained for two optimizations for sinusoidal (a,b) and spline-shaped channel walls (c,d).}
    \label{fig:learning-curves}
\end{figure*}

In general, the existence of multiple local minima of $J$, which can be obtained for significantly different sets of parameters $\mathbf{b}$ cannot be precluded. In the applied BO algorithm, the risk of stagnation of the solution near the local minimum of $J$, not being the global one, is minimized by balancing the exploration and exploitation strategies~\cite{blanchard2021output, brochu2010tutorial}. To verify the robustness of the BO algorithm and check the possible impact of the initialization procedure on the final solution, the optimization is repeated twice with different initializations. This approach allows verifying whether the BO procedure converges to the same results in terms of $J$ and $\mathbf{b}$. 
Figure~\ref{fig:learning-curves} presents the learning curves displaying the values of $J_m$ (black dots) and the current minimum of $J$ (red line) for the sinusoidal (a,b) and spline (c,d) shaped walls. The initial datasets are shaded by a semi-transparent vertical blue bar. {\color{black}Guided by the acquisition function (see Equation \ref{equ:acquisition}), the design space is sampled and evaluated. As reflected by the distribution of the tested $J$ values, samples near the learning curve correspond to the exploitation process of BO, while those with higher $J$ values are likely a result of the exploration phase.} 
For the sinusoidal wall type (Fig. \ref{fig:learning-curves}a,b), the solution converges to the optimal one already in approximately 150 iterations regardless of the initial datasets. In both cases, the optimization procedure yields an identical wall shape, characterized by $J=0.066$. For the spline-shaped wall, the situation is slightly different. Starting from the first dataset (Fig. \ref{fig:learning-curves}c), the algorithm finds the optimal solution in approximately 600 iterations, whereas 
the optimization starting from the second dataset (Fig. \ref{fig:learning-curves}d) converges earlier, in about 400 iterations. This discrepancy may be attributed to the wider input parameter space, which likely presents a greater challenge for the BO procedure in efficiently navigating toward the optimal solution. 
However, noticing the impact of the initial datasets on the convergence rate the important observation is that for both of them, the BO algorithm identifies the same optimal configuration with $J=0.1256$. This value is significantly lower than that for the channel with flat or grooved walls, but at the same time, it is almost twice as large as the value obtained for the sinusoidal wall type. To investigate the reasons for this difference, we focus on selected wall shapes and the mutual relations between $\Delta p$ and $\sigma_T$.

\begin{figure*}[h!] 
    \centering
    \includegraphics[width=0.99\linewidth]{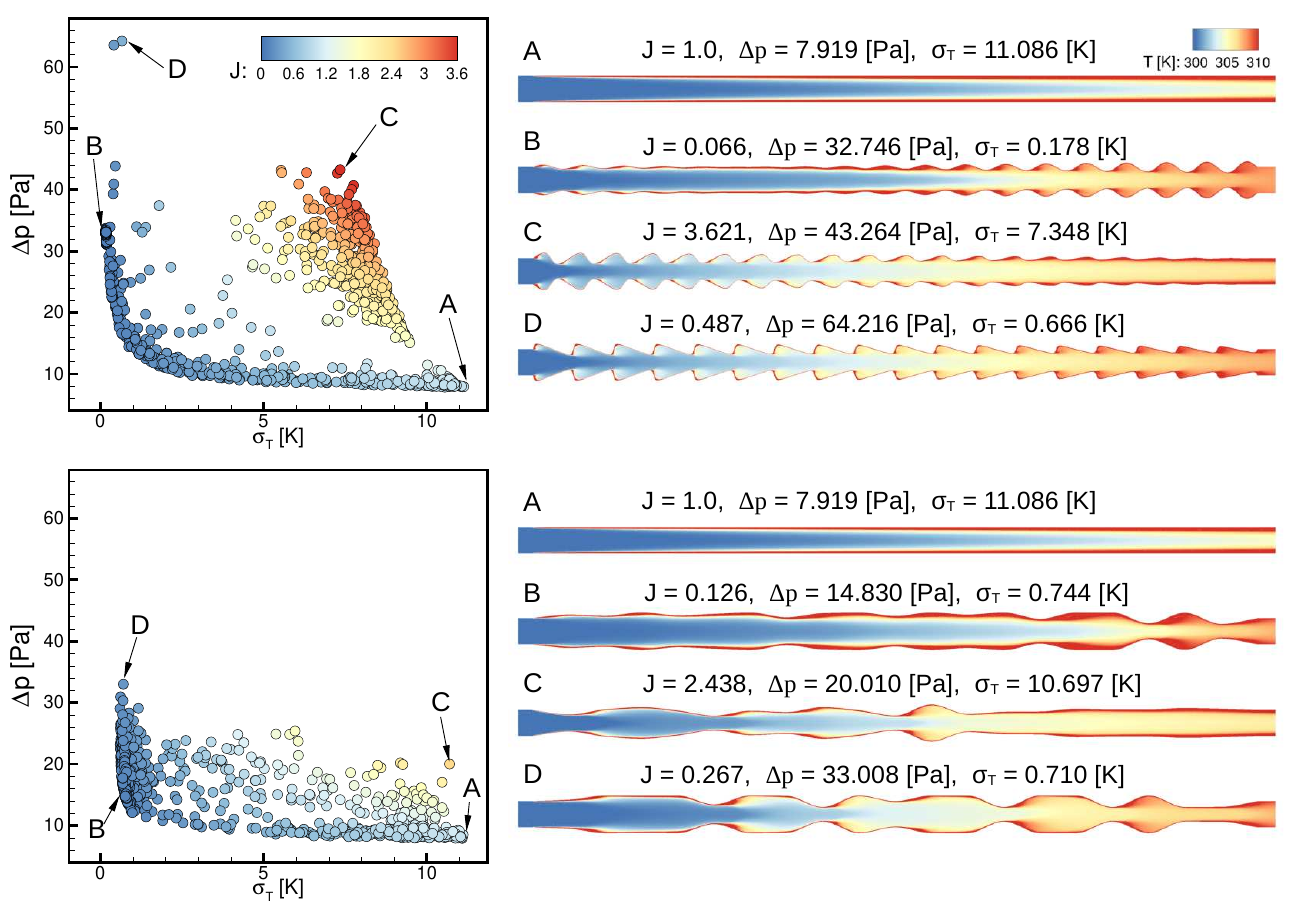}
    \caption{Scatter plot of $\Delta p(\sigma_T)$ colored by $J$ for the sinusoidal wall type (top row) and spline-shaped wall (bottom row). 
    }
    
    \label{fig:scatterplots}
\end{figure*}

Figure~\ref{fig:scatterplots} shows scatter plots of $\Delta p(\sigma_T)$ colored by $J$ for the sinusoidal wall type (top row) and spline-shaped wall (bottom row). Additionally, the wall shapes and contour maps of temperature are presented for the flat channel configuration denoted by A, the optimal configuration B, the worst configuration with the highest value of $J$, i.e. C, and the configuration yielding the highest pressure drop D. 
It can be seen that for the sinusoidal wall type, a well-defined Pareto front forms, along which the Pareto optimal configurations are located. They are characterized by the fact that no further improvement in $\sigma_T$ is possible without worsening the solution in terms of $\Delta p$. Conversely, when moving along the Pareto front, selecting a configuration with a lower $\Delta p$ results in a higher $\sigma_T$. 
Evaluating the scatter plot for the spline-shaped wall, a Pareto front is also formed, though in this case, it is not as sharp and well-defined. {\color{black}This is due to the sparsity of data within the high-dimensional design space.}
Additionally, it should be noted that the maximum $\Delta p$ for this wall type is half that of the sinusoidal wall. This is attributed to the maximum number of potential quasi-periods, which can be defined for particular shaping functions. The spline function spanned on 21 points allows for 10 periods. For the sinusoidal wall type, it has been assumed that the maximum allowable number of periods is $N=20$ and the worst configuration D is characterized by 18 periods with relatively large amplitudes. This intensifies the mixing process, leading to small $\sigma_T$ but also causes local flow separation and pressure drop, resulting in large overall $\Delta p$.
These issues will be investigated in depth in the subsequent part of the paper.

An interesting trend emerges when evaluating the impact of the wall shapes on $J$. 
The optimal configurations (cases B) for both the sinusoidal and spline-shaped walls feature a relative smoothness in the initial part of the heated area, followed by a consistent growth in the amplitude waviness. The superiority of this type of configuration can be explained as follows. Initially, when the cold fluid enters the heated section the temperature gradient between the fluid and the wall is large. This causes an intense heat transfer without the necessity of intense mixing. In effect, the pressure drop in this flow part is low while $\sigma_T$ decreases. Further downstream, as the near-wall thermal layer thickens, increasing temperature uniformity across the channel requires intensified mixing. This is achieved by an increasing waviness amplitude. It can be said that the BO procedure leads to wall shapes whose superiority can be explained based on physical principles and intuition.
Unlike the optimal configurations, the worst ones (denoted C) manifest a large deformation at the beginning of the heated section, which then diminishes downstream. In these cases, the initial waviness causes a large temperature growth in the waviness cavities that is accompanied by a significant pressure drop. The lack of wall deformation near the end of the heated section results in poor mixing and in effect, the high-temperature fluid in the near wall region is not mixed with the flow in the channel axis. The configurations yielding the highest $\Delta p$ (cases D)   are characterized by a rather large waviness along the heated wall section. This is beneficial for decreasing $\sigma_T$, but the increase in $\Delta p$ offsets the gains achieved. 

\begin{figure*}[h!] 
    \centering
    \includegraphics[width=0.99\linewidth]{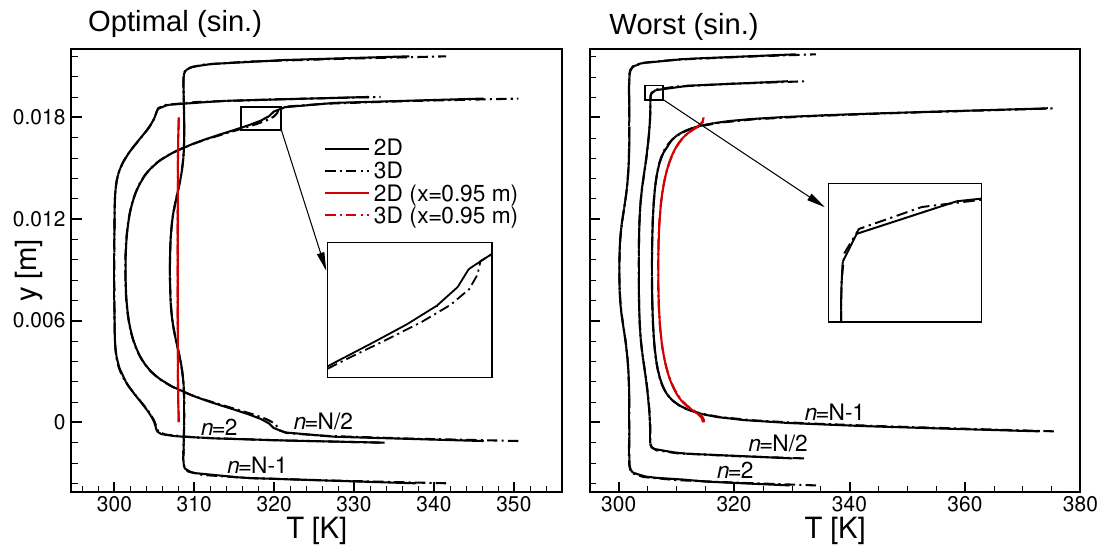}
    \caption{Profiles of the temperature in the 2nd ($n=2$), middle ($n=N/2$), and penultimate ($n=N-1$) waviness periods and at  $x=0.95$ m, for the optimal and worst configurations for the sinus-type wall.}
    \label{fig:2d3d}
\end{figure*}

\begin{figure*}[h!] 
    \centering
    \includegraphics[width=0.99\linewidth]{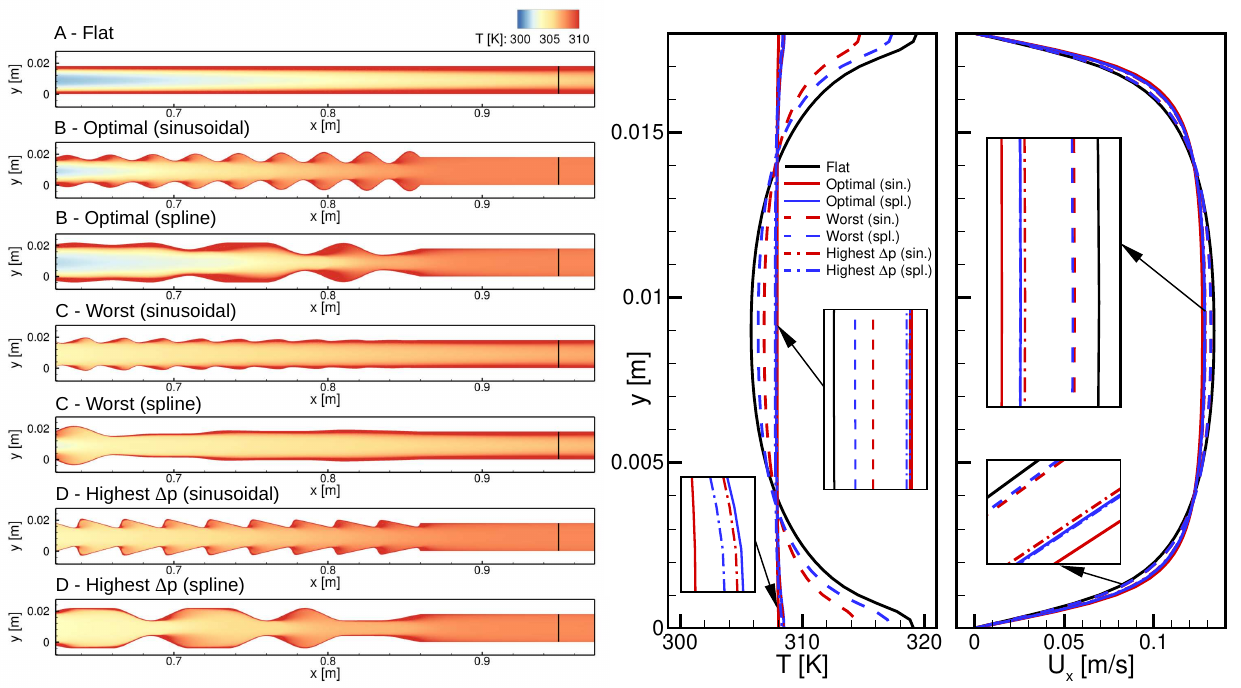}
    \caption{Contour maps of temperature for the optimal, worst, and exhibiting highest $\Delta p$ cases of the wavy wall created with sinusoidal and spline functions. Profiles of the streamwise velocity and temperature taken at $x=0.95$ m.}
    \label{fig:downstream-profiles}
\end{figure*}

{\color{black}
\subsubsection{2D vs. 3D results}\label{sec:2d3d}

As discussed at the beginning of this section, preliminary computations conducted for several 3D channel geometries revealed a predominantly two-dimensional flow character. This observation allowed the RANS model validation and optimization to be carried out using 2D geometries. The main advantage of this simplifying assumption was the significant reduction in computational time (1-2 hours for 3D on 96 CPUs vs. 5-10 minutes for 2D on 16 CPUs) - a particularly important factor in the optimization process, which required performing 900 and 1300 consecutive simulations for the sinus-type and spline-shaped walls, respectively. This saving resulted from the number of cells in the computational meshes. As already mentioned, the 2D partially unstructured meshes consisted of approximately $3.3$–$3.5 \times 10^5$ cells. In the 3D cases, the meshes were created for known selected geometries and could be manually designed as fully structured. Maintaining the condition $y^+ < 1$, and using $\Delta x$ similar to the one in 2D with $\Delta z \approx 2\Delta x$, resulted in the meshes containing $2.04 \times 10^7$ cells. To verify whether the 2D solutions obtained during the optimization process accurately capture the flow dynamics in the full 3D channels, Fig.~\ref{fig:2d3d} compares the results from the 2D and 3D cases for the optimal and worst sinusoidal wall configurations. The temperature profiles shown in this figure were extracted at the locations of maximum waviness amplitude in the 2nd ($n=2$), middle ($n=N/2$), and penultimate ($n=N-1$) waviness periods, as well as at a far downstream location at $x=0.95$~m. It can be seen that at all examined distances, the agreement between the 2D and 3D results is very good. Although minor discrepancies are present, they are negligible and barely perceptible without significant magnification. These differences are confined to the wavy channel sections near the upper and lower walls, where strong temperature gradients occur. It is also worth noting that, for the cases with spline-shaped walls, the level of disagreement between the 2D and 3D results is similarly minimal. Therefore, it can be concluded that the optimization procedure based on 2D cases is reliable and yields meaningful results for the full 3D channel configurations.


}

\subsubsection{Flow characteristics}

This section investigates how the previously outlined configurations affect the flow field across the channel and along its axis. In particular, first, we analyze to what extent the differences observed at the end of the waviness persist in the downstream straight channel part. Figure \ref{fig:downstream-profiles} presents the profiles of the streamwise velocity and temperature at the location $x=0.95$~m. 
marked with a black line in the temperature contour maps. Examining the temperature profiles reveals that both worst configurations (cases D - grey dashed lines) display a trend similar to that observed for the flat channel (case A - black solid line). In the near-wall area, the temperature is high and quickly decreases towards the central flow part. Conversely, in the optimal configurations (cases B - blue lines) and those yielding the highest $\Delta p$ (cases  C - red lines), the temperature is nearly uniform and almost identical. In cases B, the profile for the sinusoidal wall type is flatter, which is particularly evident in the near-wall area. This is also reflected by reduced $\sigma_T = 0.178$ K compared to the spline-shaped wall, where $\sigma_T = 0.744$ K. However, it should be noted that despite the relatively large difference in $\sigma_T$ (over 4 times), the actual perceived differences are small.

Evaluating the streamwise velocity profiles, it can be observed that they are very similar for all configurations and the small differences are not due to temperature variations, as the fluid properties (density, viscosity) are constant. They originate only from the dynamics of the flow inside the wavy section. Assuming that a more turbulent flow characterizes a more rectangular velocity profile, the flow in the cases B can be said to be the most turbulent. Conversely, the velocity profile for the flat channel (case A) can be treated as the least turbulent as it manifests the highest velocity in the channel axis and the lowest velocity in the near-wall region. Such a tendency is not surprising and could be expected knowing that the turbulence intensity is the main factor responsible for the mixing process.

\begin{figure}[h!] 
    \centering
    \includegraphics[width=0.99\linewidth]{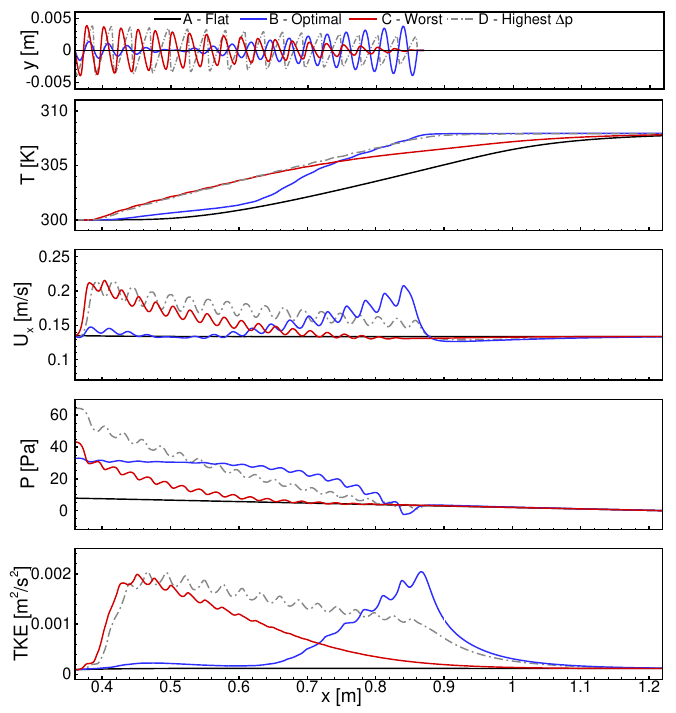}
    \caption{Axial profiles of the temperature, streamwise velocity, pressure, and turbulent kinetic energy for the cases with optimal, worst, and leading to the highest $\Delta p$ for the wavy wall created with the sinusoidal function. }
    \label{fig:axial-profiles}
\end{figure}

\begin{figure}[h!] 
    \centering
    \includegraphics[width=0.99\linewidth]{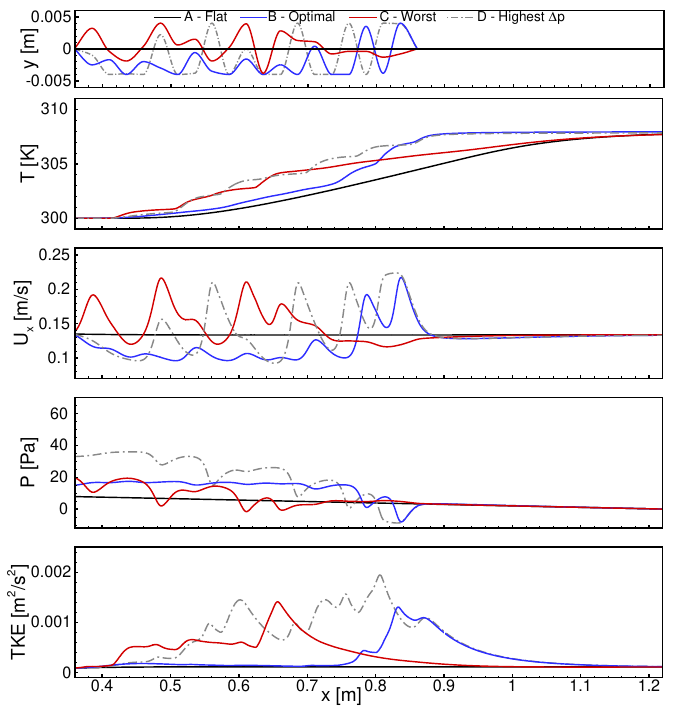}
    \caption{Axial profiles of the temperature, streamwise velocity, pressure, and turbulent kinetic energy for the cases with optimal, worst, and leading to the highest $\Delta p$ for the wavy wall created with the spline function.}
    \label{fig:axial-profiles-spline}
\end{figure}

Figures~\ref{fig:axial-profiles} and \ref{fig:axial-profiles-spline} present the axial profiles of temperature, streamwise velocity, pressure, and turbulent kinetic energy for cases A-D. Additionally, the upper subfigures depict the shapes of the lower walls in particular configurations.
Examining the temperature profiles, as expected, the slowest temperature growth is found for case A. Conversely, the configurations with the highest $\Delta p$ (cases D - dash-dotted lines) show rapid temperature growth in the initial part of the wavy wall, which then slows down in the region where the amplitude of the waviness decreases (see Fig.~\ref{fig:scatterplots}). The optimal configurations (B - solid lines) exhibit a relatively slow temperature growth in the initial part (small waviness amplitude, see Fig.~\ref{fig:scatterplots}), which then accelerates in the region where the amplitude increases. Worth noticing is that in configurations B and D the temperature at the end of the heated section reaches the same level, which almost does not change downstream. 
The streamwise velocity profiles are evidently related to the wall shapes. Compared to the flat channel, the mass conservation law enforces flow acceleration in the narrower parts of the channel and deceleration where the channel widens. Worth noting is that for the optimal sinusoidal and flat channel configurations, the velocity values up to $x=0.6$~m are similar. The pronounced differences are seen downstream when the waviness amplitude grows. Contrary, in the case of the optimal spline-shaped wall, the velocity is lower than in the flat channel over most of the heated section and it suddenly increases only near its end.     
The variation of the velocity directly translates to the local pressure distributions, which decrease/increase where the flow accelerates/decelerates. Note, that in some places it is lower than in the flat channel, e.g., in configuration D for spline-shaped wall where the velocity is large due to the narrowing of the channel.  For both optimal configurations, the pressure profiles remain relatively smooth and constant throughout the heated section. This contrasts with configuration D of the sinusoidal wall type, where the initial part of the wavy section induces the most dynamic pressure fluctuations. In this case, we attribute the sudden changes in pressure values to the waviness tilt, particularly its left-side orientation. This is the main difference between configurations D and C, where the waviness amplitude is similar but the tilt is in the opposite direction. We found that the large pressure fluctuations in configuration D are caused by local flow separation zones that form behind the steep slopes of the waviness. 

From the perspective of mixing enhancement, the values of turbulent kinetic energy (TKE) are of particular interest. They represent the magnitude of velocity fluctuations and, consequently, turbulent mixing, which plays a key role in transporting heat from the wall and redistributing it across the flow. For the optimal configurations, the waviness in the regions $0.36\leq x\leq 0.65$~m for the sinusoidal wall type and $0.36\leq x\leq 0.75$~m for the spline-shaped wall does not result in a significant increase in TKE and its level is similar to that of the flat wall. In effect, in this region, the conductive heat transfer plays a dominant role, leading to a slow temperature growth along the channel. The situation changes in the ending parts of the waviness, where TKE significantly rises and its high level causes an immediate temperature increase. As will be shown later, this is due to a relatively large vertical velocity component induced in the waviness cavities. 
The highest TKE in the initial waviness section is observed for configurations C and D of the sinusoidal wall. In these two cases, the TKE level rises steeply at the beginning of the waviness, reaching maximum values by the third waviness period. The fluctuations in TKE profiles, particularly for configuration D, are linked to pressure and velocity variations, resulting in intense mixing and a rapid temperature rise. The high level of TKE along the axis stems from particularly large TKE values within the waviness cavities.

%
\begin{figure*}[h!] 
    \centering
    \includegraphics[width=0.99\linewidth]{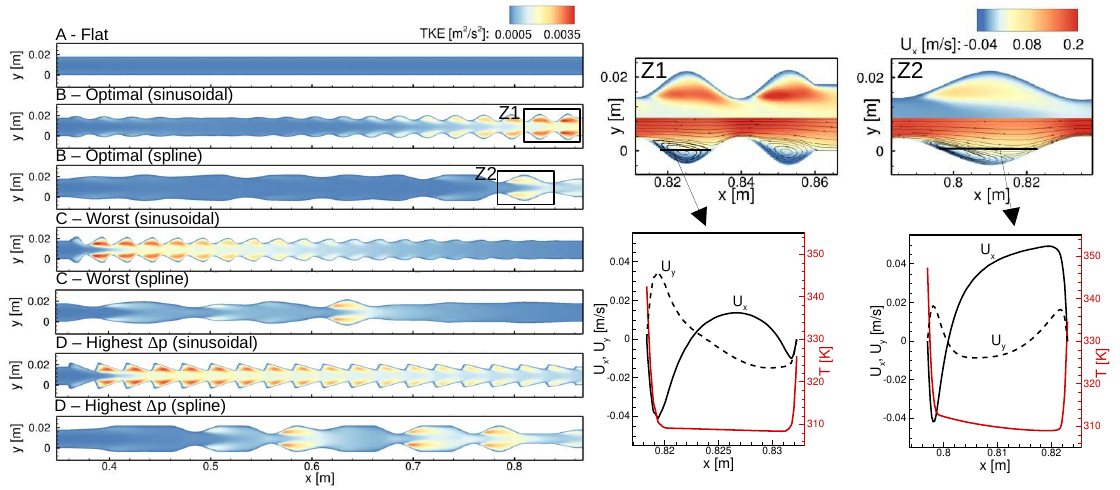}
    \caption{Contour maps of TKE for the cases with the optimal, worst, and exhibiting highest $\Delta p$ wall shapes created with sinusoidal and spline functions. Close-up views of the recirculation areas and profiles of velocity and temperature.}
    \label{fig:tke-recirc}
\end{figure*}
Contour plots of TKE are shown in Fig.~\ref{fig:tke-recirc}. Its level in the flat channel is very low and virtually constant along the channel. In cases C and D of the sinusoidal wall type, TKE distributions are characterized by the occurrence of local maxima, which start to be seen in the second waviness and are located in between the central flow part and the wall. Their repeatable distribution manifests in the fluctuating TKE profiles along the axis, as can be seen in Fig.~\ref{fig:axial-profiles}. In the configurations with spline-shaped walls, TKE visibly rises only locally behind downhill-oriented walls where the waviness amplitude is large.

In the optimal configurations, the level of TKE remains low throughout the initial parts of the heated sections, increasing predominantly only near their ends. 
In these regions, the TKE enhancement is caused by the formation of separation zones in the waviness cavities. Detailed views of the places with amplified TKE are presented in zooms labelled Z1 and Z2 to examine this phenomenon more closely. These close-ups show TKE and streamwise velocity contour maps, offering a detailed analysis of near-wall flow patterns. The occurrence of the recirculation zones within the wave troughs is manifested by a negative velocity and closed oval-like streamlines (black lines). In the sinusoidal wall configuration, these zones span most of the waviness period, whereas in the case of the spline-shaped wall, they are smaller and narrowed to the downhill waviness part. 
This narrowing is well seen in figures showing the streamwise velocity profile extracted from the middle of the waviness height. Evidently, in the case of the spline-shaped wall, the extent over which $U_x$ is negative is significantly smaller.
It is worth noting that in both configurations, the minimum values of $U_x$ are relatively large, approximately $-0.04$ m/s, which corresponds to 36\% of the inlet velocity. This results in strong recirculation and transport of fluid from the center of the waviness towards the wall. An induced vertical velocity component ($U_y$) near the wall is the primary factor contributing to the increased TKE level. Analysis of the TKE contours shows that the region of its large values begins where a high-momentum fluid (with large $U_y$) in the cavities interacts with the flow in the central part of the channel.

The robust recirculation is a prerequisite for enhanced mixing and heat exchange. Its efficiency can be measured through the temperature distribution on the walls and near its vicinity. Its profiles in both configurations exhibit clear similarities, however, in the near-wall regions, the temperature is significantly higher in the case of the spline-shaped wall. This suggests a less efficient heat transfer between the wall and fluid, translating to a less uniform temperature distribution in the cavities and the central flow part. The heat exchange efficiency between the wall and fluid also affects the wall temperature.  

\begin{figure}[h!] 
    \centering
    \includegraphics[width=0.99\linewidth]{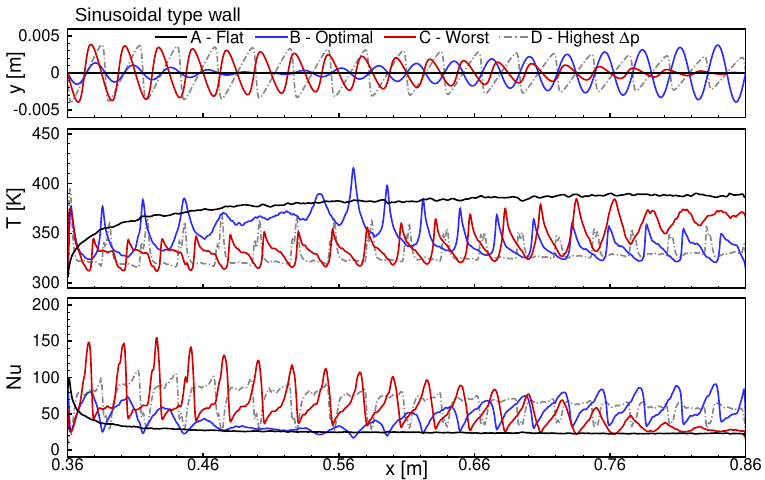}
    \caption{Profiles of temperature at the wall for optimal, worst and exhibiting highest $\Delta p$ cases of the sinusoidal type wall. The upper row indicates the shape of the particular waviness.}
    \label{fig:wall-profiles}
\end{figure}

\begin{figure}[h!] 
    \centering
    \includegraphics[width=0.99\linewidth]{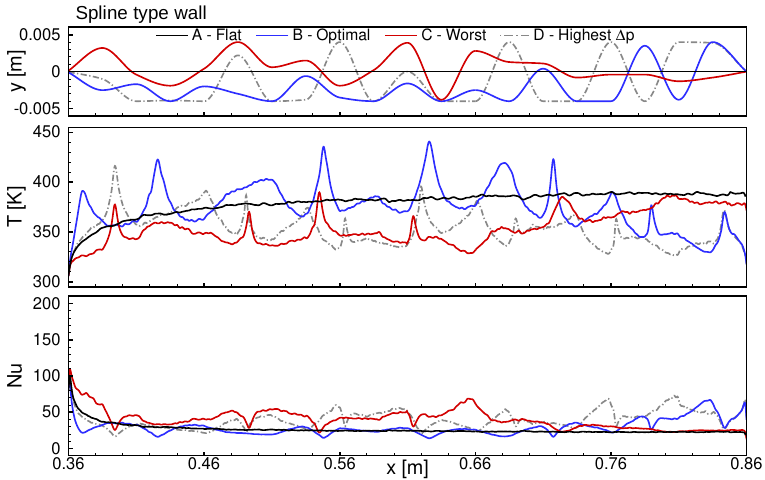}
    \caption{Profiles of the temperature at the wall for the cases with the optimal, worst, and exhibiting highest $\Delta p$ wall shapes for the spline-shaped wall. The upper row shows the shape of the particular waviness.}
    \label{fig:wall-profiles-spline}
\end{figure}

\begin{figure}[h!] 
    \centering
    \includegraphics[width=0.975\linewidth]{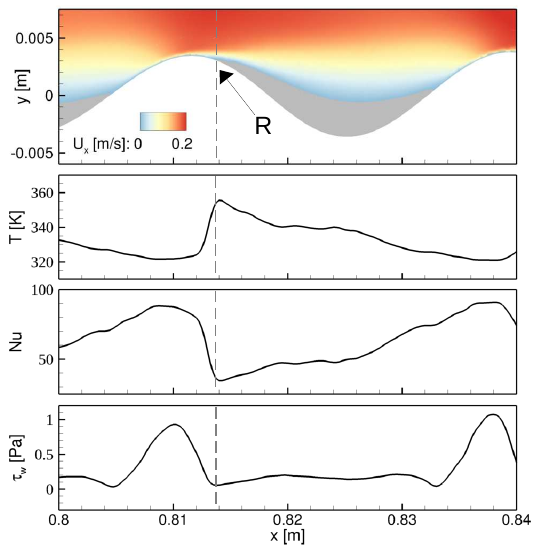}
    \caption{Contour map of $U_x$ showing a close-up of the waviness cavity, with regions where $U_x<0$, indicating separated flow, shaded in grey. Supplemented by profiles of $T$, $Nu$, and $\tau_w$. }
    \label{fig:U-T-Nu-tau}
\end{figure}

Its profiles along the wavy channel sections are presented in Fig.~\ref{fig:wall-profiles} and Fig.~\ref{fig:wall-profiles-spline} along with the distribution of the Nusselt number, $Nu=\partial T/\partial n|_w H/(T_w-T_{in})$, where $\partial T/\partial n|_w$ denotes the derivative in the wall-normal direction and $T_w$ is the wall temperature. In the case of the flat channel, initially, the wall temperature is low as the cold fluid effectively absorbs the heat due to strong convection, as can be inferred from high $Nu$ values. Then, the temperature rises smoothly, and around $x=0.75$~m reaches an almost constant level of $T\approx~388$~K, while $Nu$ decreases and approaches an asymptotic regime. A striking difference in the temperature and Nusselt number profiles obtained for wavy walls is the occurrence of local maxima and minima. 
Locations of high-temperature peaks closely match the beginnings of the recirculation zones on the downhill wall side. To investigate this issue in more detail, we analyze the velocity contours within the cavity of last waviness in the optimal configuration of the sinusoidal wall type, along with the profiles of temperature, $Nu$, and wall shear stress ($\tau_w = \mu \frac{dU}{dn}$) along the wall. These results are presented in Fig.~\ref{fig:U-T-Nu-tau}. In the velocity contour map, the grey region represents areas where $U_x < 0$. It can be observed that at the location where this region starts (marked as R), $\tau_w = 0$. In this place, the cooling effect of the wall by the flowing stream is minimal, and heat conduction (low $Nu$) becomes the key factor in heat absorption from the wall. Its relatively low efficiency causes the high wall temperature growth. Downstream of point R, $\tau_w$ rises, implying the growth of the convective heat transfer (increasing $Nu$), and in consequence, the temperature drops. This scenario takes place in every waviness cavity in all configurations. Taking into account the solutions presented in Fig.~\ref{fig:wall-profiles} it can be seen that in optimal configuration, the temperature in the first half of the heated section  ($x<0.6$~m) is larger ($Nu$ smaller) than in cases C and D. Locally, its values significantly exceed the wall temperature of the flat channel. Downstream, in the region $x>0.6$~m, the areas of elevated TKE near the wall (see Fig.~\ref{fig:tke-recirc}) increase $Nu$ and yield a reduction of the wall temperature. Its level near the end of the waviness becomes similar to that for case D for which $\sigma_T$ did not diverge substantially (see Fig.~\ref{fig:scatterplots}). On the contrary, in case C with decreasing waviness amplitude and large $\sigma_T, J$, the temperature rises and $Nu$ decreases. Analogous trends are observed in the configurations with the spline-shaped walls, see Fig.~\ref{fig:wall-profiles-spline}. The quantitative differences are observed for distributions of local temperature and $Nu$ maxima/minima, which, due to generally smoother wall shapes, occur less frequently. This factor also causes smaller longitudinal gradients of these quantities between the maxima/minima. On the other hand, it increases temperature and lowers $Nu$ values, particularly in cases C and D.    

\subsubsection{Human-assisted optimization}

Human-assisted optimization (HAO) relies on combining researchers' intuition and expertise with algorithmic or computational optimization methods. In this section, we apply this strategy with the hope that the formulated HAO-BO approach will provide a solution that at least does not diminish the superiority of the former optimal configuration, with simultaneous fulfilment of additional goals. To generally summarize the findings so far, it can be said that the configurations, which optimally balance mixing improvement (small $\sigma_T$) with pressure drop (small $\Delta p$), exhibit a distinct characteristic along the heated section. The initial heated section features a relatively smooth wall with small waviness amplitudes, leading to a low $\Delta p$. In this flow part, the initially cold fluid is effectively heated, forming a near-wall thermal layer. As this layer thickens, the temperature gradient decreases, and heat transport across the flow weakens. Thus, the necessity of mixing arises, and the BO method 'suggests' an increase in the waviness amplitude. At this point, one might ask whether these BO-guided optimal configurations can be further improved or if some of their negative aspects can be mitigated. Regarding the second issue, two problems should be considered. From an engineering perspective, manufacturing a wavy wall incurs additional costs and can be challenging, especially in cases with irregular wall shapes. Hence, the best would be to eliminate waviness in the regions where its contribution to achieving the optimization goal is weak. It can be seen in Figs.~\ref{fig:axial-profiles} and \ref{fig:axial-profiles-spline} that in the initial part of the heated sections, the temperature in the channel axis increases smoothly and not significantly faster than in the configuration with the flat wall. On the other hand, the waviness in this region causes pressure loss, and most importantly, it leads to a non-uniform wall temperature distribution with local high-temperature peaks (see Figs.~\ref{fig:wall-profiles}-\ref{fig:wall-profiles-spline}). Assuming that a given heat exchanger needs to be produced at a low cost, its walls should be thin. However, in this case, large temperature gradients would result in high thermal stresses, which, over time, could deform or damage the wall.
Hence, taking these arguments into account, the natural attempt for improvement of the BO-optimal configuration seems to be to flatten the initial channel section. Certainly, this approach should minimize the pressure drop, and simultaneously, considerably streamline the manufacturing process and reduce the overall complexity of the channel. To 
explore the consequences of such a combined intuition and engineering-driven optimization we focus on the sinusoidal type wall, which is characterized by better performance in terms of $J$. 
%
\begin{figure*}[h!] 
    \centering
    \includegraphics[width=0.99\linewidth]{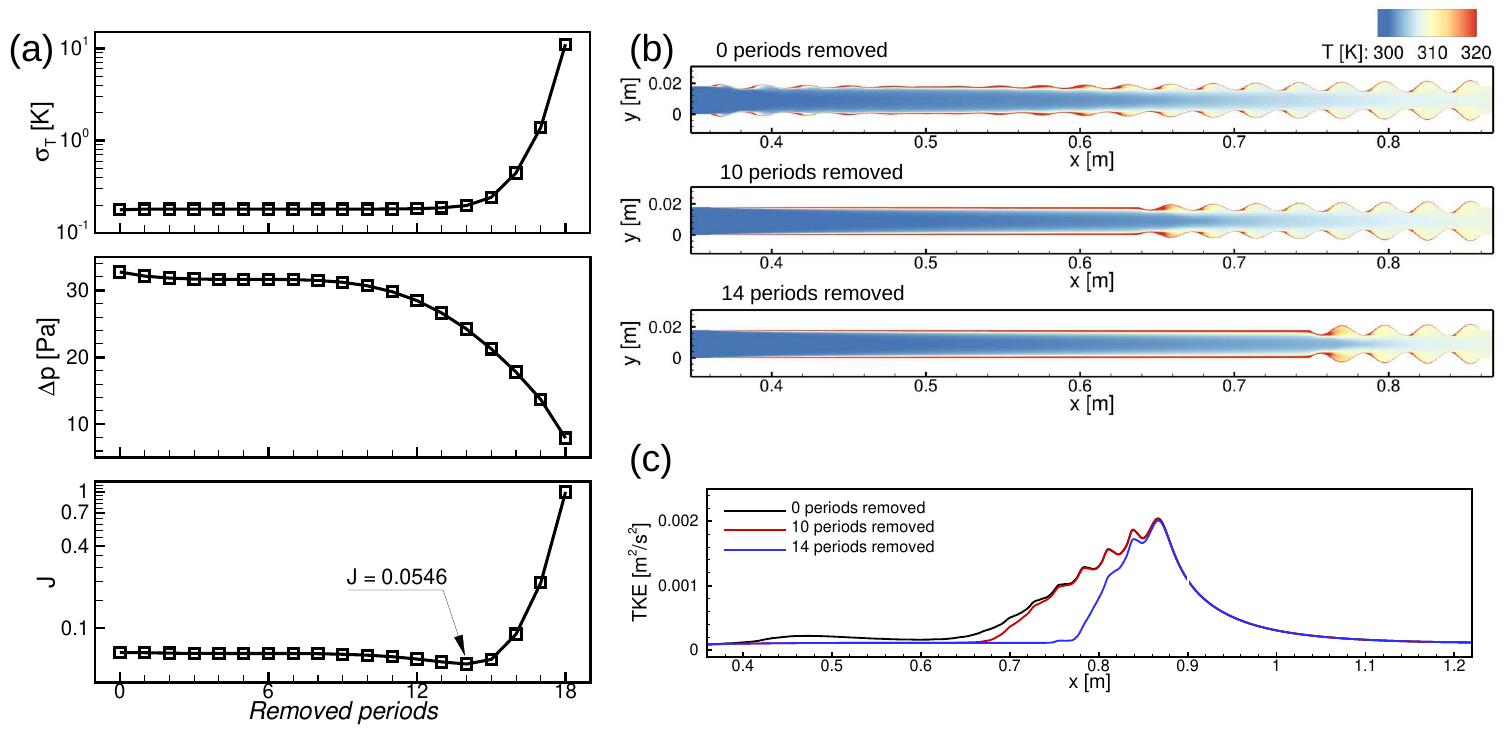}
    \caption{Profiles of $\sigma_T$, $\Delta p$, and $J$ as a function of the removed periods. ; sample contour maps of $T$ illustrating three shapes of walls (b); streamwise profiles of TKE for three considered cases taken at $y=0.009$ m, i.e. the middle section of the channel.}
    \label{fig:applicability}
\end{figure*}
%
In the performed analysis, starting from the beginning of the heated section, successive periods of waviness in the BO-optimal configuration are replaced with a flat surface. Then, RANS simulations are performed for such a new configuration and  $\sigma_T$, $\Delta p$ and $J$ are calculated.
Figure \ref{fig:applicability} shows their profiles as a function of the number of removed waviness periods. Please note that $\sigma_T$ is presented in the logarithmic scale. Additionally, the contour maps of the temperature for three selected cases with corresponding profiles of TKE along the channel axis are presented. To support the following discussion quantitatively, Table \ref{tab:data} presents the detailed values of $J$, $\Delta p$, and $\sigma_T$ for all smoothened configurations.
It can be seen that without the first wave, $\sigma_T$ rises by 1.7\% (from $0.178$~K to $0.181$~K), whereas $\Delta p$ drops by 1.9\% (from $32.746$~Pa to $32.129$~Pa), and this causes a small improvement in $J$ (from $0.0664$ to $0.0661$). Continuing the flattening process up to 10 initial waves has almost no impact on $\sigma_T$, though it does reduce $\Delta p$ slightly. 
Notable differences start to occur when the 11th waviness period is removed. From this point, $\sigma_T$ pronouncedly increases while $\Delta p$ decreases, with the latter occurring more rapidly, leading to a visible decrease in $J$. In effect, after removing the 14th waviness period, $J$ reaches the minimum at $J=0.0546$, which is 17.7\% lower than the initial value. Further flattening of waviness causes a drop in $\Delta p$, but this is accompanied by a rapid increase in $\sigma_T$, leading to a subsequent rise in $J$. Hence, the optional configuration is the one with 14 removed waves. Apart from the significant reduction in wall complexity, an interesting observation can be made regarding this case. The 4 remaining waves, constituting 22.2\% of the original 18 waves, contribute to the majority of the mixing. While not exact, this result aligns strikingly with the Pareto 80/20 principle.

The contour maps of the temperature presented in Fig. \ref{fig:applicability}b show the solutions obtained for the BO-optimal configuration and the cases with 10 and 14 periods removed. It can be seen that downstream of the waviness the temperature fields are nearly identical, which confirms the above findings regarding the impact of waviness complexity on the mixing intensity. The temperature profiles on the flattened wall parts (not shown) cover with that previously obtained for the flat channel (Cf. Fig.~\ref{fig:surface-sine}). Their distribution in the wavy part closely resembles the profile in the BO-optimal configuration.  Regarding the level of TKE (Fig. \ref{fig:applicability}c), the removal of 10 waviness periods does not bring significant changes. As expected, in the flattened section, TKE is lower compared to the BO-optimal configuration; however, the differences are generally minor. In the new optimal configuration, TKE rises steeply as the waviness begins, ultimately reaching the same level as in the two other cases by the end. It turns out that the last 4 waves are sufficient to enhance TKE, ensuring efficient mixing without a significant overall pressure drop.

\begin{table}[ht]
    \centering
    \begin{tabular}
    {p{80pt}p{35pt}p{35pt}p{35pt}}
    \hline
    \hline
        Removed periods & $J$ & $\Delta p$ [Pa] & $\sigma_T$ [K] \\ \hline
        0 (full waviness)  & 0.0664 & 32.746 & 0.178  \\
1  & 0.0661 & 32.129 & 0.181  \\
2  & 0.0656 & 31.835 & 0.181  \\
3  & 0.0653 & 31.681 & 0.181  \\
4  & 0.0653 & 31.649 & 0.181  \\
5  & 0.0652 & 31.649 & 0.181  \\
6  & 0.0652 & 31.635 & 0.181  \\
7  & 0.0652 & 31.606 & 0.181  \\
8  & 0.0650 & 31.479 & 0.181  \\
9  & 0.0644 & 31.208 & 0.181  \\
10 & 0.0634 & 30.677 & 0.181  \\
11 & 0.0618 & 29.792 & 0.182  \\
12 & 0.0594 & 28.465 & 0.183  \\
13 & 0.0566 & 26.613 & 0.187  \\
\textbf{14} & \textbf{0.0546} & \textbf{24.225} & \textbf{0.198}  \\
15 & 0.0590 & 21.260 & 0.243  \\
16 & 0.0905 & 17.799 & 0.446  \\
17 & 0.2158 & 13.687 & 1.384  \\
18 (flat channel) & 1.0000 & 7.919  & 11.086 \\
\hline
\hline
    \end{tabular}
    \caption{Values of $J$, $\Delta p$ and $\sigma_T$ in a function of the waviness periods replaced by the flat wall. The optimal solution is highlighted in bold font.}
    \label{tab:data}
\end{table}

\section{Conclusions \label{conclusions}}
{\color{black}
This paper presented an optimization study of heated wall surfaces in a channel using machine learning. The design of the wall surfaces was guided by a learning framework that combines BO with RANS simulations conducted in ANSYS Fluent code.} The primary goal of the research was to minimize temperature variations $\sigma_T$ arccos the channel at possibly the smallest pressure drop $\Delta p$. The cost function $J$ balanced these two factors and was normalized by the reference values obtained for the flat channel. Two wall deformation methods were compared: a sinusoidal function and a spline function employing the Piecewise Cubic Hermite Interpolation Polynomial (PCHIP). {\color{black}The former is characterized by four parameters, while the latter offers larger design freedom with $19$ parameters.}

{\color{black}
The higher-dimensional design of the spline-shaped wall presented greater challenges for optimization.
Both sinusoidal and spline-shaped wall optimizations were repeated twice with different initializations. For the sinusoidal wall, convergence was achieved in approximately $150$ BO iterations regardless of the initial dataset. In contrast, the convergence of the spline-shaped wall was highly dependent on initialization, requiring between $400$ and $600$ iterations. Additionally, the data sparsity associated with the 19-dimensional design space required a larger number of data points to derive the Pareto front that effectively captures the trade-off between the temperature variations $\sigma_T$ and the pressure drop $\Delta p$.

The optimized sinusoidal wall type outperformed the spline-shaped wall although the latter considered a higher-dimensional design.  The minimum objective value, $J$, achieved by the sinusoidal wall was only half of that obtained with the spline-shaped wall.
}
Specifically, the sinusoidal wall type led to significantly lower $\sigma_T$, though at the cost of a higher $\Delta p$. The enhanced performance of the sinusoidal wall likely resulted from a larger number of periods, which could be formed along the heated wall, i.e. 20 vs. 10 quasi-periods for the spline-shaped wall. This resulted in a lower effective amplitude--to--wavelength ratio ($A/\lambda$), reducing the potential for improving mixing, however, at lower pressure drops.

The optimized surfaces for both approaches featured a relatively smooth initial section, minimizing $\Delta p$, followed by a region of increased amplitude that encouraged the formation of recirculation zones, which enhanced TKE and flow mixing. The worst-performing configurations, in terms of $J$, were characterized by surface shapes that were the reverse of the optimal design. This motivated performing a human-assisted optimization (HAO) to reduce potential manufacturing costs of obtained optimal solution and minimize thermal stresses. HAO was applied only for the configuration with the sinusoidal wall type and involved successive replacement of small-amplitude waviness periods by flat segments starting from the beginning of the heated section. It has been shown that flattening up to 10 initial waves has minimal impact on $J$, yet the reduced complexity of the new configurations could facilitate manufacturability. Further removal of waviness periods led to a rapid decrease in $\Delta p$, with flattening 14 waviness periods yielding a new minimum $J$ that performed 17.7\% better than the best wall shape found during the BO procedure.

In summary, the studies demonstrated the significant potential of wavy wall surfaces in enhancing mixing and heat transfer processes. The combined use of the BO-RANS method, complemented by HAO, proved to be a highly effective optimization tool. A natural progression of this research is to apply the formulated algorithm to higher Reynolds number flows and cases with large temperature differences, as seen in combustion devices. In such studies, it would be beneficial to verify CFD results with more accurate modeling methods. For instance, using large eddy simulations, though computationally expensive, would increase the reliability of the solutions and provide deeper insights into flow dynamics. This is planned for future research.

\section{Acknowledgements}

From the Polish side, this work was supported by the National Science Centre under grant UMO-2020/39/B/ST8/01449 and statutory funds of the Czestochowa University of Technology BS/PB 1-100-3011/2024/P. We gratefully acknowledge Poland's high-performance Infrastructure PLGrid ACK Cyfronet AGH and PCSS for providing computer facilities and support within computational grants no PLG/2023/016731 and PL0356-01. From the Chinese side, this work was supported by the National Science Foundation of China (NSFC) through grant 12172109 by the Guangdong Basic and Applied Basic Research Foundation under grant 2022A1515011492, and by the Shenzhen Science and Technology Program under grants JCYJ20220531095605012, KJZD20230923115210021 and 29853MKCJ202300205. 

\bibliographystyle{ieeetr}


%


\end{document}